\documentclass[%
 reprint,
 superscriptaddress,
 nofootinbib,
 amsmath,amssymb,
 prc,
 floatfix
]{revtex4-2}

\usepackage{graphicx}
\usepackage{dcolumn}
\usepackage{bm}
\usepackage{xcolor}
\usepackage{wasysym}
\usepackage{physics}
\usepackage{amsmath}
\usepackage{subfigure}
\usepackage[pdftex,colorlinks,linkcolor = blue,urlcolor  = blue,citecolor = blue]{hyperref}

\newcommand{\be}{\begin{equation}}
\newcommand{\ee}{\end{equation}}
\newcommand{\bea}{\begin{eqnarray}}
\newcommand{\eea}{\end{eqnarray}}

\begin{document}

\preprint{APS/123-QED}

\title{Exploring the statistical properties of the neutron-deficient $^{109}$In isotope \\with the Oslo method}%

\author{M.~Markova}
\email{maria.markova@fys.uio.no}
\affiliation{Department of Physics, University of Oslo, N-0316 Oslo, Norway}
\affiliation{Norwegian Nuclear Research Centre, Norway}

\author{A.~C.~Larsen}
\email{a.c.larsen@fys.uio.no}
\affiliation{Department of Physics, University of Oslo, N-0316 Oslo, Norway}
\affiliation{Norwegian Nuclear Research Centre, Norway}

\author{P.~von Neumann-Cosel}%
\affiliation{%
 Institut f\"{u}r Kernphysik, Technische Universit\"{a}t Darmstadt, D-64289 Darmstadt, Germany
}%
\affiliation{Norwegian Nuclear Research Centre, Norway}

\author{E.~Litvinova}%
\affiliation{%
 Department of Physics, Western Michigan University, Kalamazoo, Michigan 49008, USA
}%

\affiliation{%
GANIL, CEA/DRF-CNRS/IN2P3, F-14076 Caen, France
}%

\author{S.~Goriely}
\affiliation{Institut d’Astronomie et d'Astrophysique, Université Libre de Bruxelles, CP 226, B-1050 Brussels, Belgium}

\author{L.~T.~Bell}
\affiliation{Department of Physics, University of Oslo, N-0316 Oslo, Norway}
\affiliation{Norwegian Nuclear Research Centre, Norway}

\author{T.~K.~Eriksen}
\affiliation{Department of Physics, University of Oslo, N-0316 Oslo, Norway}
\affiliation{Norwegian Nuclear Research Centre, Norway}


\author{A.~G\"{o}rgen}
\affiliation{Department of Physics, University of Oslo, N-0316 Oslo, Norway}
\affiliation{Norwegian Nuclear Research Centre, Norway}

\author{M.~Guttormsen}
\affiliation{Department of Physics, University of Oslo, N-0316 Oslo, Norway}
\affiliation{Norwegian Nuclear Research Centre, Norway}

\author{E.~F.~Matthews}
\affiliation{Department of Nuclear Engineering, University of California, Berkeley, California 94720, USA}

\author{A.~J.~Nordberg}
\affiliation{Department of Physics, University of Oslo, N-0316 Oslo, Norway}

\author{W.~Paulsen}
\affiliation{Department of Physics, University of Oslo, N-0316 Oslo, Norway}
\affiliation{Norwegian Nuclear Research Centre, Norway}

\author{L.~G.~Pedersen}
\affiliation{Department of Physics, University of Oslo, N-0316 Oslo, Norway}

\author{F.~Pogliano}
\affiliation{Department of Physics, University of Oslo, N-0316 Oslo, Norway}
\affiliation{Expert Analytics AS, N-0179 Oslo, Norway}

\author{E.~Sahin}
\affiliation{Department of Physics, University of Oslo, N-0316 Oslo, Norway}
\affiliation{Norwegian Nuclear Research Centre, Norway}

\author{S.~Siem}
\affiliation{Department of Physics, University of Oslo, N-0316 Oslo, Norway}
\affiliation{Norwegian Nuclear Research Centre, Norway}

\author{T.~G.~Tornyi}
\affiliation{HUN-REN Institute for Nuclear Research (ATOMKI), H-4026 Debrecen, Hungary}

\date{\today}

\begin{abstract}

The nuclear level density (NLD) and the $\gamma$-ray strength function (GSF) of the neutron-deficient $^{109}$In isotope were extracted for the first time with data from the $^{106}$Cd$(\alpha,p\gamma)$$^{109}$In reaction using a combination of the Oslo and the shape methods. Both quantities are consistent with those of neighboring Cd and Sn nuclei, but show substantial discrepancies with currently available model predictions. In contrast to earlier observations in the neighboring isotopic chains, $^{109}$In does not exhibit any significant enhancement of the dipole strength near the neutron separation energy. To interpret this feature, random-phase time-blocking approximation calculations have been performed for $^{109}$In and the neighboring $^{110,112}$Sn nuclei. The experimental data were also employed to estimate cross sections and rates of the radiative neutron- and proton-capture reactions, $^{108}$In($n,\gamma)$$^{109}$In and $^{108}$Cd($p,\gamma)$$^{109}$In, respectively, with the reaction code TALYS. Our ($p,\gamma)$ cross section is in excellent agreement with direct measurements over a wide range of proton energies, while the ($n,\gamma)$ cross section demonstrates notable deviations from predictions in the JINA REACLIB library. The new results on the statistical properties of $^{109}$In provide valuable constraints that may help address the problem of large model uncertainties compromising the accuracy of astrophysical $p$-process simulations.

\end{abstract}

\maketitle
\section{\label{sec 1: introduction}Introduction}

Over the past two decades, extensive theoretical and experimental efforts have been dedicated to advance our understanding of the electric dipole ($E1$) response of nuclei, particularly dipole excitations at relatively low energies around and below the neutron emission threshold. Such excitations form an enhancement of the $E1$ strength superimposed on the low-energy tail of the Isovector Giant Dipole Resonance (IVGDR), and are commonly referred to as the Pygmy Dipole Resonance (PDR) \cite{Savran2013, Bracco2019, Lanza2022}. Within the most widespread, rather simplified picture, the PDR strength emerges due to out-of-phase oscillations of a neutron excess versus an isospin-saturated core. This perspective, however, as well as the degree of collectivity attained in PDR states, has been challenged in numerous theoretical studies \cite{Roca-Maza2012, Reinhard2013, Papakonstantinou2014, Lanza2022} and continues to be scrutinized with novel experimental approaches \cite{Spieker2020, Weinert2021, Neumann-Cosel2024}. 

In addition to the research of the microscopic content of PDR states \cite{Spieker2020, Weinert2021} and the interplay between its isovector and isoscalar components \cite{Endres2012, Pellegri2014, Savran2018}, mapping the evolution of the PDR in different isotopic chains remains a highly relevant scientific objective. The potential connection between the neutron excess, or skin, and the emergence of the PDR strength has directed the focus of these studies toward heavy, neutron-rich isotopes \cite{Adrich2005, Rossi2013, Wieland2018}. This was additionally motivated by the expected impact of the enhanced $E1$ strength around the neutron separation energy on radiative neutron-capture rates of interest for the astrophysical $r$ (rapid neutron-capture) process, highlighted for the first time in Ref.~\cite{Goriely98}. At the same time, the neutron-deficient region of the nuclear chart remains comparatively unexplored.
Heavy, neutron-deficient nuclei, accessible with currently available experimental techniques, still lie well in the vicinity of the valley of stability and are therefore not suitable for exploring the theoretically suggested PDR driven by proton oscillations \cite{Tsoneva2008}. Nevertheless, even a few neutrons away from stability, these nuclei may provide important insight into the development of the low-lying $E1$ strength for cases with a low neutron-to-proton asymmetry. 


The Sn mass region, and the Sn isotopic chain in particular, has been thoroughly studied in the past decade with a variety of theoretical and experimental techniques to investigate the evolution of the low-lying $E1$ strength, commonly associated with the PDR, as a function of increasing neutron excess. A clear increase of the summed strength extracted in nuclear resonance fluorescence experiments on the $^{112,116,120,124}$Sn isotopes \cite{Govaert98, Ozel-Tashenov2014, Muscher2020} was discussed in Ref.~\cite{Muscher2020}. A more complete $E1$ response provided by a combination of particle-$\gamma$ coincidence, Coulomb excitation, and photo-dissociation data in Ref.~\cite{Markova2024} does not reveal any apparent increase of the low-lying $E1$ strength from $^{111}$Sn to $^{124}$Sn. Instead, the integrated strength remains rather constant throughout the whole chain of the studied nuclei. Similar analyses of Cd \cite{Larsen2013_Cd} and Pd \cite{Eriksen2014} isotopes indicate a weak enhancement of the low-lying $E1$ strength toward heavier nuclei; making any strong conclusions based on these results is, however, complicated due to the large experimental uncertainties. For unstable, neutron-rich $^{130,132}$Sn, strengths extracted from Coulomb excitation in inverse kinematics were reported to indicate a significant PDR-like enhancement above the neutron threshold \cite{Adrich2005, Klimkiewicz2007}, consistent with the expected increase of the PDR strength with increasing neutron-to-proton asymmetry. However, no information about the strength below the neutron threshold is available for these cases. It remains an open question, however, whether the PDR strength is reduced in neutron-deficient nuclei and, if so, to what extent it is reduced compared to previously studied neutron-rich cases. In addition to being essential for a deeper understanding of the PDR, experimental data on such nuclei may offer critical constraints for microscopic and phenomenological models applied in large-scale astrophysical simulations.

In this regard, the Oslo method \cite{Larsen2011} is an excellent tool for studying such statistical properties of nuclei as the $\gamma$-ray strength function (GSF), characterizing the average $\gamma$-transition probabilities, and the nuclear level density (NLD). Both of these characteristics are the key inputs in Hauser-Feshbach statistical model \cite{Hauser1952} calculations of radiative proton and neutron capture rates relevant for heavy-element nucleosysntesis. Research in this field has been done by the Oslo group primarily for nuclei of interest for the $s$ (slow neutron-capture) and $i$ (intermediate neutron-capture) processes (see e.g. \cite{CrespoCampo2016, Pogliano2022}). Two stable isotopes, $^{89}$Y \cite{Larsen2016} and $^{92}$Mo \cite{Tveten2016}, have also been studied in the context of the $p$ process. In the Sn mass region, an attempt to approach $\beta^+$/EC-radioactive nuclei was made by studying the short-lived $^{105}$Cd \cite{Larsen2016} and $^{111}$Sn \cite{Markova2023}. However, these nuclei did not display any significantly different structural properties compared with the neighboring stable isotopes. In general, the low-lying $E1$ strength in even-$Z$ nuclei in this mass range is quite well studied, whereas data on odd-$Z$ nuclei remain rather scarce. 

In this paper, we continue exploring the low-lying $E1$ strength in the above-mentioned mass region by investigating the unstable, neutron-deficient isotope $^{109}$In. Both the NLD and the GSF were extracted from $p-\gamma$ coincidence data using the Oslo method. The details of the experimental procedure and the method are outlined in Sec.~\ref{sec 2: Method and Experiment}. The results are compared with available theoretical models and experimental data for the neighboring isotopes and discussed within a broader context of the systematics of the low-lying $E1$ strength in the Sn mass region (Sec.~\ref{sec 3: Results}). Moreover, the newly extracted NLD and GSF were used to calculate data-constrained $(n,\gamma)$ and $(p,\gamma)$ cross sections and rates, potentially offering new insights relevant for $p$ process simulations (Sec.~\ref{sec 4: Astro}). Conclusions and an outlook are provided in Sec.~\ref{sec 5: Conclusion}.                        
\section{\label{sec 2: Method and Experiment}Experiment and methodology}

\subsection{\label{subsec 2.1: Experimental details}Experimental details and the Oslo method}
The experiment was performed at the Oslo Cyclotron Laboratory (OCL) in 2021. A 96.7\%-enriched, 1.03~mg/cm$^2$-thick $^{106}$Cd target was used to investigate the $^{109}$In isotope through the $^{106}$Cd($\alpha, p \gamma$)$^{109}$In reaction. A 23-MeV $\alpha$ beam impinging on the target with an average intensity of 4 nA was produced by the MC-35 Scanditronix cyclotron. An additional 1-hour run on a 1~mg/cm$^2$-thick natural $^{12}$C target was performed to calibrate $\gamma$-ray spectra. Approximately $1.3\times10^{9}$ $p-\gamma$ coincidence events were recorded after 4 full days of beam time.

This experiment was among the first efforts to access unstable, neutron-rich or neutron-deficient odd-$Z$ isotopes from stable targets through the ($\alpha, p$) reaction at the OCL. The configuration of the setup used in this work was similar to that employed in the most recent studies by the Oslo group (see, e.g., Ref.~\cite{Pogliano2022}). The target was placed in the center of a spherical chamber, $\approx$16.3 cm away from 30 large-volume 3.5$^{\prime\prime}\times$ 8$^{\prime\prime}$ LaBr$_{3}$(Ce) crystals mounted on a icosahedron-shaped frame, forming the Oslo SCintillator ARray (OSCAR) \cite{Zeiser2020}. The array has a total efficiency of $\approx$57\% and an energy resolution of $\approx2.2$\% at $E_{\gamma} = 1332$ keV. The time resolution in terms of the full width at half maximum of the prompt $p-\gamma$ coincidence peak achieved in this experiment was $\approx 3$ ns. The detectors were calibrated using Doppler-corrected $\gamma$-ray energies corresponding to transitions from the first excited states in $^{15}$N (average $E_{\gamma}\approx 5283$ keV) and $^{19}$F (1236 keV and 2583 keV) produced in the ($\alpha,p$) reaction on the oxidized calibration target $^{12}$C, as well as the annihilation peak.

\begin{figure*}[t]
\includegraphics[width=1.0\linewidth]{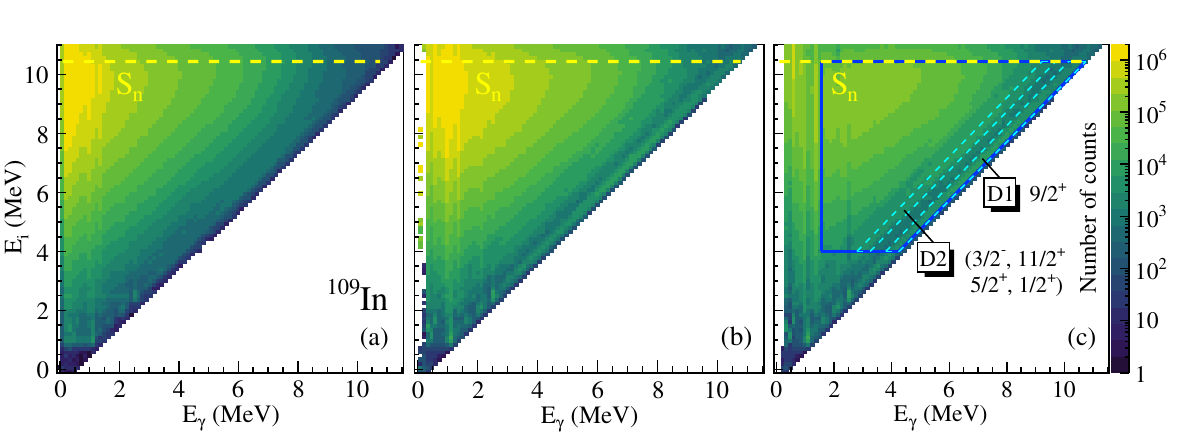}
\caption{\label{fig 1: Matrices}
(a) Experimental $p-\gamma$ coincidence data, (b) unfolded, and (c) primary matrices for $^{109}$In. Yellow dashed lines mark the neutron separation energy. The ranges of excitation and $\gamma$-ray energies used for the analysis with the Oslo method are shown with solid blue lines in (c). The gates on the ground state with spin-parity $9/2^+$ (D1) and a cluster of states with spin-parities $3/2^-$, $11/2^+$, $5/2^+$, $1/2^+$ (D2) used for the shape method are shown with cyan dashed lines in (c). Excitation-energy and $\gamma$-ray-energy bins are 124-keV wide.}
\end{figure*}

Particles were detected by a Si $\Delta E-E$ telescope array SiRi \cite{Guttormsen2011}, located in backward position and covering polar angles from 126$^{\circ}$ to 140$^{\circ}$. The array consists of eight trapezoidal telescopes with $\approx$130-$\mu$m-thick $\Delta E$  and $1550$-$\mu$m-thick $E$ counters. The former chips are additionally segmented into eight strips covering 2$^{\circ}$ each. Protons produced in the ($\alpha, p$) reaction were separated from other outgoing particles using the $\Delta E-E$ technique, and their energies deposited in the $E$ and $\Delta E$ detectors at different angles were converted into the excitation energies of the residual nucleus $^{109}$In for each event. The elastic peak and a cluster of the four excited states below the excitation energy of 1.2 MeV in the ($\alpha,p$) channel, as well as a cluster of the ground and the first excited states in the ($\alpha,d$) channel were used for a linear calibration of the SiRi detectors. The excitation energy resolution achieved in this experiment, estimated for the highest energy protons in the ($\alpha, p$) channel, is approximately 250~keV. The $p-\gamma$ coincidence matrix containing $\gamma$-ray spectra extracted for initial excitation energies $E_i$ of $^{109}$In up to the neutron threshold energy, $S_n$($^{109}$In)= 10.439 MeV, is shown in Fig.~\ref{fig 1: Matrices}(a). 

All $\gamma$-ray spectra in the $p-\gamma$ coincidence matrix were unfolded using the OSCAR detector response function \cite{Zeiser2020}, simulated with the \textsc{Geant4} software package \cite{Agostinelli2003, Allison2006, Allison2016}, and an iterative technique outlined in Ref.~\cite{Guttormsen1996}. The so-called Compton subtraction method (see Ref.~\cite{Guttormsen1996} for more details) was implemented in the unfolding procedure to preserve the statistical fluctuations of the observed $\gamma$-ray spectra in the unfolded spectra, shown in Fig.~\ref{fig 1: Matrices}(b).

The Oslo method employed in this work utilizes the proportionality of the primary (first-generation) $\gamma$-ray distribution to the statistical properties of interest, i.e., the NLD and the GSF. Hence, the next step of the analysis involves the extraction of primary transitions from the $\gamma$-ray cascades contained in the unfolded matrix. For this, an iterative subtraction technique called the \textit{first generation method} was applied \cite{Guttormsen1987}. This procedure is based on the subtraction of a weighted sum of lower-lying unfolded spectra (containing second-, third-, and higher-generation transitions) from an unfolded spectrum at a certain excitation energy, where the weighting factors are directly proportional to the primary $\gamma$-ray distribution. Further details of this method can be found in, e.g., Ref.~\cite{Larsen2011}. The primary matrix for $^{109}$In produced with this technique is shown in Fig.~\ref{fig 1: Matrices}(c). Due to the assumption that $\gamma$ decay from an excited state is independent of whether the state was populated directly in the reaction or via a $\gamma$ cascade from higher-lying states, the applicability of the first-generation method is limited to rather high excitation energies. For the further analysis, the range of excitation energies was therefore restricted to energies $4$~MeV $\leq E_i \leq S_n(^{109}$In), where $\gamma$ decay is expected to be preceded by the formation of the compound nucleus (see Ref.~\cite{Larsen2011} for a more detailed discussion). Under- and over-subtraction of some low-energy $\gamma$ transitions imposes a lower limit for $\gamma$-ray energies at $\approx$1.6 MeV. The quasi-continuum region of the primary matrix used in the subsequent analysis is indicated by blue solid lines in Fig.~\ref{fig 1: Matrices}(c). Some traces of $\gamma$-rays from the de-excitation of the first $2^{+}$ state in $^{108}$Cd produced by proton emission from $^{109}$In are seen at excitation energies above $\approx 9$ MeV in the primary $p-\gamma$ coincidence matrix. This, however, does not affect the analysis of $^{109}$In in any significant way due to the chosen lower limit imposed on the $\gamma$-ray energy.

The extracted primary $\gamma$-ray distributions $P(E_{\gamma},E_i)$ are further decomposed in terms of the $\gamma$-transmission coefficient $\mathcal{T}_{i\rightarrow f}$ (proportional to the GSF) and the NLD $\rho_f$ as
\begin{equation}
\label{eq:1}
    P(E_{\gamma},E_i)\propto \mathcal{T}_{i\rightarrow f}\rho_f.
\end{equation}
Here, $P(E_{\gamma},E_i)$ corresponds to the probability of $\gamma$ decay from an excited state $i$, or rather from a group of excited states in the excitation-energy bin $E_i$, to the lower lying states $f$ (excitation-energy bin at $E_f$) by emitting a photon of energy $E_{\gamma}=E_i-E_f$. The decomposition was done following the procedure outlined in Ref.~\cite{Schiller2000}, yielding the functional forms of $\rho_f=\rho(E_f)=\rho(E_i-E_{\gamma})$ and $\mathcal{T}_{i\rightarrow f}=\mathcal{T}(E_{\gamma})$. The latter is assumed to be a function of $\gamma$-ray energy only in accordance with the generalized Brink-Axel hypothesis\footnote{The $\gamma$-transmission coefficient, as well as the $\gamma$-ray strength function proportional to it, is assumed to be independent of the energies, spins, and parities of the initial and final states, and to depend only on the energy of the $\gamma$ transition between them.} \cite{Brink1955, Axel1962}. This is one of the central assumptions of the Oslo method, and the extent to which it holds in practice has been addressed in several previous studies (see, e.g., \cite{Guttormsen2016, Campo2018}) and, in particular, for the neighboring Sn isotopes in Refs.~\cite{Markova2021, Markova2022}. We henceforth assume that it also holds well for the case of $^{109}$In. 

Equation (\ref{eq:1}) is grounded in Fermi’s golden rule and the Hauser-Feshbach theory of statistical reactions (as demonstrated in Refs.~\cite{MIDTBO_1} and Ref.~\cite{MIDTBO_2}) and is expected to hold for compound excited states in the chosen region of the primary matrix, similarly to the previous analyses in this mass region \cite{Larsen2013_Cd, Eriksen2014, Pogliano2022, Markova2024}.

As outlined in Ref.~\cite{Schiller2000}, the decomposition provides the functional forms of the solutions for the NLD and the $\gamma$-transmission coefficient defined in the general form as:

\begin{equation}
\label{eq:2}
\begin{split}
    \tilde{\rho}(E_i-E_{\gamma}) = &A\rho(E_i-E_{\gamma})\exp[\alpha (E_i-E_{\gamma})],\\
    \tilde{\mathcal{T}}(E_\gamma) = &B\mathcal{T}(E_\gamma)\exp[\alpha E_{\gamma}].
\end{split}
\end{equation}
Here, $\rho(E_i-E_{\gamma})$ and $\mathcal{T}(E_\gamma)$ are determined from a $\chi^2$-fit of the experimental primary matrix $P(E_{\gamma},E_i)$ with the following relation \cite{Schiller2000, Larsen2011}:

\begin{equation}
\label{eq:3}
P_{th}(E_{\gamma}, E_i)=\frac{\mathcal{T}(E_{\gamma})\rho(E_i-E_{\gamma})}{\sum_{E_{\gamma}=E_{\gamma}^{min}}^{E_i}\mathcal{T}(E_{\gamma})\rho(E_i-E_{\gamma})}, 
\end{equation}
where $P_{th}(E_{\gamma}, E_i)$ is the theoretical distribution of primary $\gamma$ rays.

With one set of solutions found, $\rho(E_i-E_{\gamma})$ and $\mathcal{T}(E_\gamma)$, an infinite set of equivalent solutions $\tilde{\rho}(E_i-E_{\gamma})$ and $\tilde{\mathcal{T}}(E_\gamma)$ can be constructed by applying the scaling factors $A$ and $B$ and modifying the slope of the resulting functions with the parameter $\alpha$. To single out the unique physical solutions for the NLD and the $\gamma$-transmission coefficient, these parameters must be constrained using additional experimental data. 

\subsection{\label{subsec 2.2: Normalization}Standard Oslo method procedures for the normalization of the NLD and GSF}

For the majority of Oslo method analyses applied to stable nuclei, the parameters $A$, $B$, and $\alpha$ are determined using information on low-lying discrete levels, available in the ENSDF compilation \cite{ensdf}, together with $s$- or $p$-wave neutron resonance data listed in the \textit{Reference Input Parameter Library} \cite{Capote2009} and the \textit{Atlas of Neutron Resonances} \cite{Mughabghab18}. This part of the analysis is described in great detail in earlier publications, e.~g., in \cite{Larsen2011, Pogliano2022, Markova2024}. In this work, the normalization of the $^{109}$In data uses this standard procedure as a starting point, and we therefore briefly summarize it in this section. The additional procedures employed in this work to extract the NLD and the GSF of $^{109}$In are provided in the following section.

To constrain the scaling parameter $A$, the NLD from Eq.~(\ref{eq:2}) is anchored to the low-lying discrete levels in the excitation energy range where the level scheme is considered complete. Another anchor point, which determines the slope of the $\rho(E_x)$ function, is the NLD at the neutron separation energy, $\rho(S_n)$, obtained from the experimental value of the $s$-wave (and/or $p$-wave) neutron resonance spacing, $D_0$ ($D_1$, respectively). For a target nucleus with spin-parity $J_t^{\pi_t}$ of the ground state, $D_0$ can be expressed using the density of levels accessed in $s$-wave neutron capture ($J_t\neq 0$ for this example):
\begin{equation}
\label{eq:4}
    D_0=\frac{1}{\rho(S_n, J_t+1/2, \pi_t)+\rho(S_n, J_t-1/2, \pi_t)}.
\end{equation}
Assuming parity equipartition at $S_n$ (supported by data \cite{Lokitz2004, Kalmykov2007} and microscopic NLD models \cite{Goriely2008, Hilaire2012}), the parity dependence is removed and substituted with a factor of $1/2$. The spin-dependence of the NLD is handled by introducing the spin distribution $g(E_x, J)$ from Refs.~\cite{Ericson58, Gilbert65}
\begin{equation}
\label{eq:5}
    g(E_x, J) \simeq \frac{2J+1}{2\sigma^2(E_x)}\exp\left[-\frac{(J+1/2)^2}{2\sigma^2(E_x)}\right],
\end{equation}
and expressing the \textit{partial} (spin-dependent) NLD $\rho(E_x, J)$ through the \textit{total} NLD $\rho(E_x)$ as
\begin{equation}
\label{eq:6}
    \rho(E_x, J)=\rho(E_x)g(E_x, J).
\end{equation}
The spin-cutoff parameter $\sigma(E_x)$ in Eq.~(\ref{eq:5}) is often adopted from Ref.~\cite{Egidy05} (based on the relation for the rigid-body moment of inertia, RMI for short) as
\begin{equation}
\label{eq:7}
    \sigma_{\text{RMI}}^2(E_x)=0.0146A^{5/3}\frac{1+\sqrt{1+4a(S_n-E_1)}}{2a},
\end{equation}
where $a$ and $E_1$ are the level-density parameter and the back-shift parameter for the back-shifted Fermi gas (BSFG) model from Ref.~\cite{Egidy05}. This form of the spin-cutoff was used in some of the earlier works of the Oslo group, e.~g., in Refs.~\cite{Larsen2013_Cd, Guttormsen2022}. In some cases, external experimental data suggest that the RMI regime might not be fully attained at $S_n$ (see discussion in \cite{Markova2023} and references therein), and Eq.~(\ref{eq:6}) is reduced with an additional factor, as was done, for example, in  Ref.~\cite{Larsen2023}. Alternatively, the spin-cutoff model from  Ref.~\cite{Gilbert65} (G\&C for short), providing roughly a 20\% reduction of the RMI value, is chosen:
\begin{equation}
\label{eq:8}
\sigma_{\text{G\&C}}^2(S_n)=0.0888a\sqrt{\frac{S_n-E_1}{a}}A^{2/3}.
\end{equation}
This approach was adopted, for example, in Ref.~\cite{Markova2024}. In other cases, a combination of both models is used for the normalization of the NLD (see Ref.~\cite{Pogliano2022}). It is clear that the choice of $\sigma(S_n)$ represents a potential source of systematic uncertainty in the normalization procedure and should be accounted for in the total uncertainty. The specific choice of the spin-cutoff parameter for the $^{109}$In case is discussed in the following section.

Consolidating Eqs.~(\ref{eq:4})-(\ref{eq:6}), the NLD at the neutron threshold can be expressed as:
\begin{equation}
\label{eq:9}
\rho(S_n)=\frac{1}{D_0}\frac{2\sigma^2(S_n)}{(J_t+1)\exp(-\frac{(J_t+1)^2}{2\sigma^2(S_n)})+J_t\exp(-\frac{J_t^2}{2\sigma^2(S_n)})}.
\end{equation}
A similar equation can be derived for the $p$-wave neutron resonance spacing $D_1$ by taking into account the corresponding states populated through the capture of neutrons with orbital angular momentum $\ell=1$.

The lower $E_{\gamma}$ limit, imposed by the first-generation procedure failing to subtract some of the low-energy $\gamma$ transitions, restricts the excitation-energy range for the extracted NLD to energies $\approx 1-2$ MeV below $S_n$. The data must therefore be interpolated across this gap up to the $\rho(S_n)$ value. This is commonly done with the use of either the BSFG (Eq.~(3) in Ref.~\cite{Egidy05}) or the constant temperature (CT) \cite{Ericson60, Gilbert65} model. For relatively small interpolation gaps, the exact choice of the model typically has little impact on the final results. It is interesting to note that most nuclei studied by the Oslo group over the past years exhibit a distinct CT trend at excitation energies below their respective neutron threshold values \cite{Guttormsen2015}. Specifics of the interpolation model used for $^{109}$In are discussed in the next section.

These procedures allow us to determine the scaling $A$ and the slope $\alpha$ of the NLD, where the latter also gives the slope of the GSF. As a final step, the scaling $B$ of the GSF can be extracted using the average total radiative width $\langle \Gamma_{\gamma}\rangle$ from $s$- and/or $p$-wave neutron capture experiments \cite{Mughabghab18, Capote2009}. The expression for the average total radiative width $\langle \Gamma_{\gamma}(E_x, J, \pi)\rangle$
from Ref.~\cite{Kopecky1990} [Eq.~(3.1)] can be adapted to account for the states accessed through the capture of $\ell=0$ or $\ell=1$ neutrons. For example, for the $s$-wave case and a target nucleus with $J_t\neq 0$, the GSF $f(E_{\gamma})$ can be expressed through the value of $\langle \Gamma_{\gamma}\rangle$ as \cite{Larsen2011}:
\begin{align}
\label{eq:10}   
    \langle\Gamma_{\gamma}\rangle=&\langle\Gamma(S_n, J_t\pm1/2, \pi_t)\rangle=\frac{1}{2\rho(S_n, J_t\pm1/2, \pi_t)} \notag\\&\times\int_{E_{\gamma}=0}^{S_n}dE_{\gamma} E_{\gamma}^3f(E_\gamma)\rho(S_n-E_{\gamma})\notag\\&\times\sum_{J=-1}^{1}g(S_n-E_{\gamma},J_t\pm1/2+J).
\end{align}
Here, we used the proportionality of the $\gamma$-transmission coefficient $\mathcal{T}(E_{\gamma})$ and the GSF $f(E_{\gamma})$, $\mathcal{T}(E_{\gamma})=2\pi E_{\gamma}^3f(E_{\gamma})$, for dipole transitions expected to dominate in the studied energy region \cite{Kopecky1990}. Within the Oslo method, it is not possible to distinguish between magnetic ($M$) and electric ($E$) types of transitions, unless external experimental information on the $M1$ distribution is available. The GSF extracted with Eq.~(\ref{eq:9}) is therefore the sum of $E1$ and $M1$ dipole strengths. The GSF is extrapolated to low (and in some cases high) $E_{\gamma}$ energies to calculate the integral in Eq.~(\ref{eq:9}) as discussed in Ref.~\cite{Larsen2011}.

The excitation-energy dependence of the spin cutoff entering the spin distribution in Eq.~(\ref{eq:9}) is commonly adopted from Ref.~\cite{Capote2009}:
\begin{equation}
\label{eq:11}  
    \sigma^2(E_x) = \sigma_d^2 + \frac{E_x-E_d}{S_n-E_d}[\sigma^2(S_n)-\sigma_d^2],
\end{equation}
where $\sigma_d$ is the value of the spin cutoff that can be extracted from tabulated discrete levels at excitation energy $E_d$. 

\subsection{\label{subsec 2.3: Normalization 109In}Normalization of the NLD and GSF in $^{109}$In}

With neutron resonance data at hand, the normalization of both the NLD and the GSF becomes a relatively straightforward process. However, they are not available for some stable and unstable nuclei close to the valley of stability, including $^{109}$In, due to the lack of suitable targets for neutron-capture experiments. To tackle the normalization of the $^{109}$In data, some additional steps, outlined in the following subsections, are required. 

\subsubsection{Extracting the normalization coefficients for $^{109}$In from the systematics}

In some cases, systematics for neighboring nuclei can be used to constrain the values of $\rho(S_n)$ and $\langle \Gamma_{\gamma}\rangle$ \cite{Larsen2013_Cd, Markova2024}. For odd-$Z$ nuclei, however, this approach is somewhat challenging, as only a limited number of stable isotopes are available to establish reliable systematics. This problem was, for example, encountered in the recent analysis of $^{127}$Sb \cite{Pogliano2022}, which relied largely on neutron resonance data from the Sn and Te isotopic chains. Likewise, very little information is available on In isotopes (except for the stable $^{113}$In and the long-lived $^{115}$In), that could potentially be used in the normalization procedure for the neutron-deficient $^{109}$In. 

\begin{figure}[t]
\includegraphics[width=1.0\linewidth]{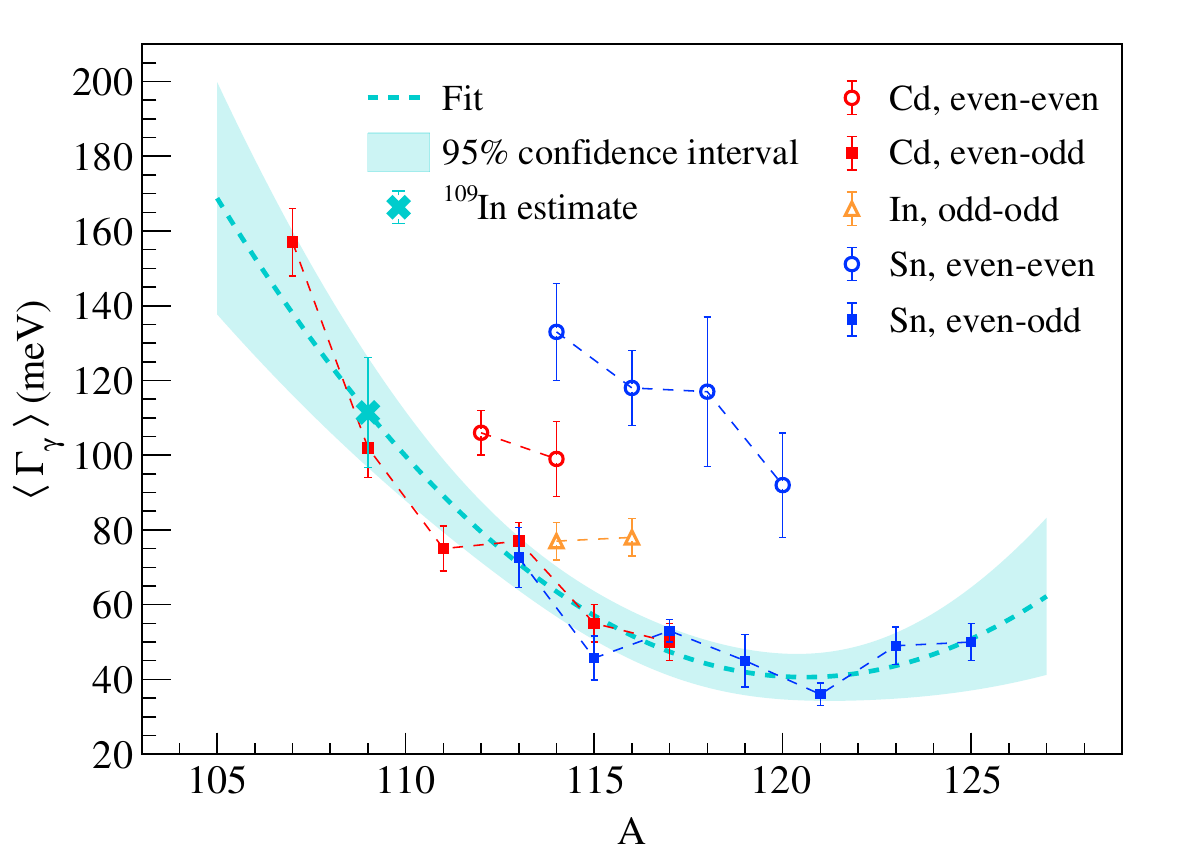}
\caption{\label{fig 2: Systematics Gg}
Experimental systematics of the average total radiative widths $\langle\Gamma_{\gamma}\rangle$ for Cd, In, and Sn isotopes. The experimental values are taken from Ref.~\cite{Mughabghab18}. The quadratic fit to the experimental data for the odd Cd and Sn isotopes and the corresponding 95\% confidence interval are shown as a light-blue dashed line and a light-blue shaded area. The $\langle\Gamma_{\gamma}\rangle$ value used for the normalization of the $^{109}$In data, extracted from the fit, is indicated with a light-blue cross.
}
\end{figure}

For the In isotopes, located one proton below the $Z=50$ closure, an approach similar to that adopted for $^{127}$Sb is suggested by the abundant neutron resonance data available for Sn and Cd nuclei. The $\langle \Gamma_{\gamma}\rangle$ value for the odd-even $^{109}$In was therefore extracted from s-wave $\langle\Gamma_{\gamma}\rangle$ values for the even-odd Sn and Cd isotopes from Ref.~\cite{Mughabghab18}. In the present case, a $\chi^2$ fit with a quadratic function was applied to the data for even-odd Sn and Cd nuclei plotted as a function of mass number (see Fig.~\ref{fig 2: Systematics Gg}). The quadratic fit was chosen as it provides a significantly better fit to the experimental data than a linear function (in contrast to the case in Ref.~\cite{Pogliano2022}), while maintaining a minimal set of assumptions made regarding the observed trend. All values are shown together with the quadratic fit in Fig.~\ref{fig 2: Systematics Gg}.

An analogous extraction of $\rho(S_n)$ \cite{Larsen2013_Cd, Markova2022, Pogliano2022} suggests a comparatively broad span of values, due to differing trends in the systematics for the Sn and Cd isotopes. For $^{109}$In, the predicted value of $\rho(S_n)$ is approximately $10^7$~MeV$^{-1}$ based on the Sn systematics, and about $10^6$~MeV$^{-1}$ based on the Cd systematics. This sets the potential systematic uncertainty in the slope of the NLD (and, consequently, the GSF) to at least $70-80\%$.

\subsubsection{Constraining the GSF slope with the shape method}

To reduce the uncertainty in the slope shared by the NLD and the GSF, the so-called \textit{shape method} was proposed in Ref.~\cite{Wiedeking2020}. This technique has been successfully applied in different mass regions \cite{Wiedeking2020, Guttormsen2022, Markova2022, Mucher2023} to reconstruct the general shape of the GSF (and, therefore, the slope of the NLD) from data on the direct decay to low-lying states. The number of counts $N$ associated with these decays is confined within a diagonal ($D$) in the primary matrix. The number of counts for a given $E_i$ bin can be written in a general form as \cite{Wiedeking2020}:
\begin{equation}
\label{eq:12} 
    N_D(E_i, J_f)\propto f(E_{\gamma})E^3_{\gamma}\sum_{J_f}\sum_{J_i=J_f-1}^{J_i=J_f+1}\sigma(E_i, J_i)g(E_i, J_i),
\end{equation}
where $f(E_{\gamma})E_{\gamma}^3$ is proportional to the $\gamma$-transmission coefficient, $\sigma(E_i, J_i)$ is the population cross-section for a state with spin $J_i$ at excitation energy $E_i$, and $g(E_i, J_i)$ is the intrinsic spin distribution given by Eq.~(\ref{eq:5}).  In cases when a diagonal contains more than one state, the first sum runs over all spins $J_f$ of the final states. As in the Oslo method, parity equipartition is adopted in Eq.~(\ref{eq:12}), and only dipole transitions are taken into account. This restricts the applicability of the shape method to excitation energies where dipole transitions are dominant. As a next step, the population cross section is assumed to be independent of spin $J_i$ in the chosen region of excitation energies $E_i$. For dipole decays from a set of initial states to the final states with spins $J_f$, confined within the diagonal $D$ at the excitation-energy bin $E_i$ in the primary matrix, the GSF can therefore be expressed as:
\begin{equation}
\label{eq:13} 
    f(E_{\gamma})\propto\frac{N_{D}(E_i, J_f)}{E_{\gamma}^3\sum_{J_f}\sum_{J_i=J_f-1}^{J_i=J_f+1} g(E_i, J_i)}.
\end{equation}
Considering decays within two separate diagonals $D_1$ and $D_2$ from the excitation-energy bin $E_i$, one can construct a ratio between two values of the GSF at different $\gamma$-ray energies:
\begin{equation}
\label{eq:13} 
    \frac{f(E_{\gamma 1})}{f(E_{\gamma 2})}=\frac{N_{D1}(E_i, J_f^{D1})}{N_{D2}(E_i, J_f^{D2})}\frac{E_{\gamma 2}^3\sum_{J_f^{D2}}\sum_{J_i^{\prime}=J_f^{D2}-1}^{J_i^{\prime}=J_f^{D2}+1} g(E_i, J_i^{\prime})}{E_{\gamma 1}^3\sum_{J_f^{D1}}\sum_{J_i=J_f^{D1}-1}^{J_i=J_f^{D1}+1} g(E_i, J_i)},
\end{equation}
with $E_{\gamma 1}=E_i-E_f^{D1}$ and $E_{\gamma 2}=E_i-E_f^{D2}$, $E_f^{D1}$ and $E_f^{D2}$ being the final excitation energies of the states in the diagonals. 

The spin-dependence of the cross section term was addressed and partly accounted for in Ref.~\cite{Guttormsen2022}. The fraction of the experimentally covered and the intrinsic spin distributions, proportional to this cross section, was estimated for the ($p,p^{\prime}\gamma$) and $(d,p\gamma)$ reactions and included in the GSF ratios in Eq.~(\ref{eq:13}). A higher average angular momentum transfer is expected in the present case of the ($\alpha, p$) reaction (approximately $10\hbar$, according to a semi-classical estimate). The populated spin distribution is therefore assumed to follow approximately the intrinsic one, and no additional corrections were made.

The ratios provided by Eq.~(\ref{eq:13}) are extracted within an excitation-energy range in the quasi-continuum up to the neutron threshold. Finally, the pairs of GSF values at different $\gamma$-ray energies are connected together with the so-called sewing technique outlined in Ref.~\cite{Wiedeking2020}. In this way, the shape of the GSF (in arbitrary units) is reconstructed from $\gamma$ transitions feeding both the first and the second diagonals. The absolute values of the extracted GSFs can be further obtained by scaling them to the available data in the vicinity of the neutron threshold.

To constrain the slopes of the NLD and the GSF of the $^{109}$In isotope, we use two diagonals shown in Fig.~\ref{fig 1: Matrices}(c): the first diagonal corresponding to the direct decay to the ground state with $J^{\pi}$=9/2$^+$ and the second corresponding to a cluster of higher-lying states with $J^{\pi}=3/2^-$, $11/2^+$, $5/2^+$, $1/2^+$. The position of the gates applied to these diagonals was optimized to minimize the potential contribution from other neighboring states. The first excited $1/2^-$ state at $\approx 650$~keV is only weakly populated through the decay of higher-lying states and cannot be reliably used for the shape method. The excited states above $\approx 1335$ keV are also not suitable for this procedure primarily due to the unknown spin assignment of some states. An analogous case in which the shape method was applied to diagonals containing several discrete states is also described in Ref.~\cite{Wiedeking2020}.

\subsubsection{Consolidating the shape method and the systematics for the normalization of $^{109}$In}

To proceed with the extraction of the GSF with the shape method and to use it for the normalization of the Oslo method strength, the spin-cutoff parameter at the neutron separation energy is required. The $\sigma(S_n)$ value enters the excitation-energy dependence of the spin-cutoff provided by Eq.~(\ref{eq:11}) and can significantly alter the slope of the resulting GSFs. 

As mentioned in Sec.~\ref{subsec 2.2: Normalization}, Eq.(\ref{eq:7}) or Eq.(\ref{eq:8}) are usually adopted in the Oslo method analysis together with Eq.(\ref{eq:11}) to model the spin-cutoff parameter dependence on the excitation energy. In the present case, however, the parameterizations of the level density parameter $a$ and the energy back-shift $E_1$ from Refs.~\cite{Egidy05, Egidy2009}, commonly used in Eq.(\ref{eq:7}) and Eq.(\ref{eq:8}), cannot be reliably applied. The functional form of the NLD extracted from the decomposition in Eq.~(\ref{eq:1}) exhibits a clear BSFG trend, however, the $a$ and $E_1$ parameters from Refs.~\cite{Egidy05, Egidy2009} fail to reproduce the shape of the NLD across the wide range of slopes suggested by the Cd and Sn isotopes. This is an additional complication that does not allow us to use these parameters to estimate the $\sigma(S_n)$ value in a well-justified manner.

\begin{figure}[t]
\includegraphics[width=1.0\linewidth]{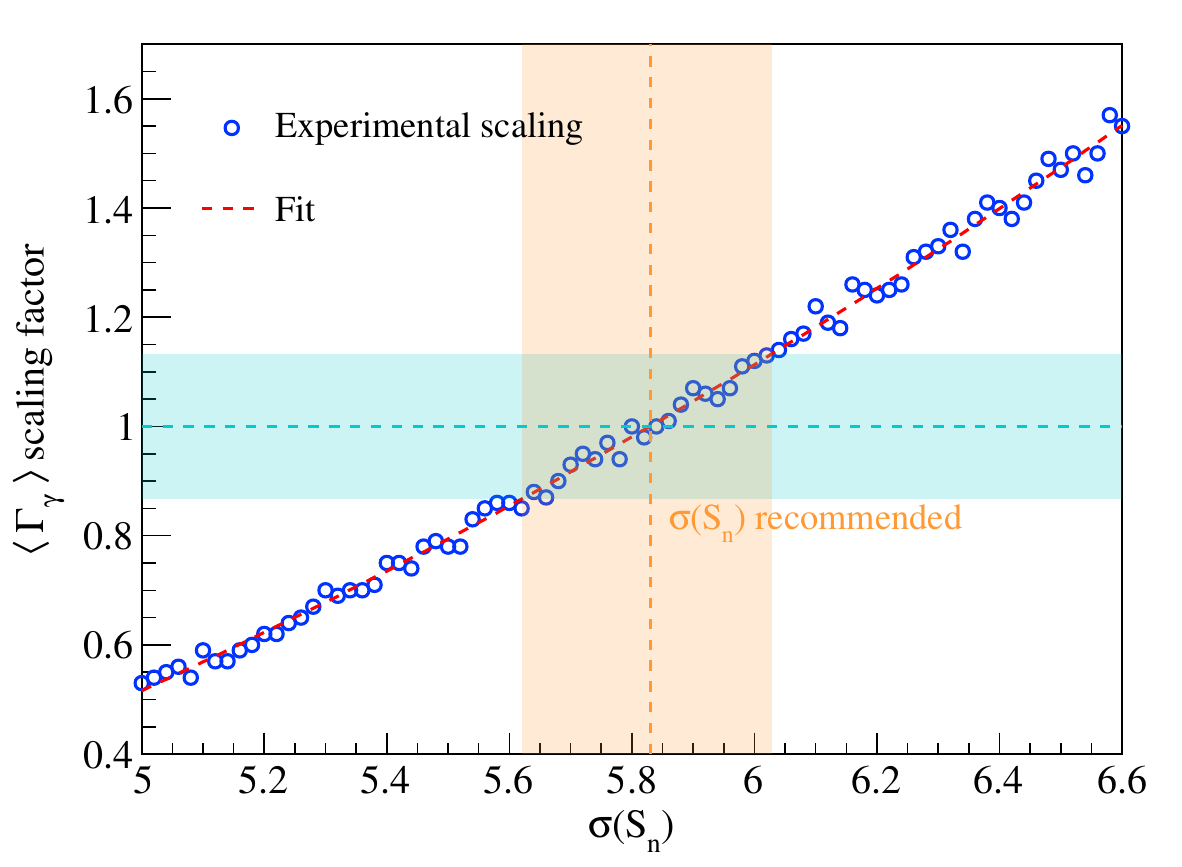}
\caption{\label{fig 3: Gg reduction factor} Scaling factor applied to $\langle \Gamma_{\gamma}\rangle$ versus the spin-cutoff parameter. A quadratic fit to the data is shown with the dashed red line. The horizontal light-blue area corresponds to the 95\% confidence interval for the $\langle \Gamma_{\gamma}\rangle$ value from the systematics, shown also in Fig.~\ref{fig 2: Systematics Gg}. The vertical orange area shows the corresponding range of spin-cutoff values.}
\end{figure}

As a consequence, the NLD and the spin-cutoff parameter at $S_n$, as well as the level density parameters $a$ and $E_1$, required for the normalization of the data with the Oslo and the shape methods are essentially unknown. To tackle this problem and to proceed with the normalization, we assume the following. Firstly, the shape of the NLD follows the BSFG model at relatively high excitation energies, and we can, thus, search for the optimal parameters $a$ and $E_1$ reproducing this shape. Secondly, we adopt the $\langle \Gamma_{\gamma}\rangle$ for $^{109}$In from the systematics for the Cd and Sn nuclei shown in Fig.~\ref{fig 2: Systematics Gg}. Thirdly, we employ the shape method to constrain the slope of the NLD and the GSF. The problem is, therefore, reduced to searching for the spin-cutoff parameter $\sigma(S_n)$ to be used in both the Oslo and the shape methods, which is also consistent with the optimal parameters $a$ and $E_1$. As mentioned above, this value affects the slope of the shape method GSFs, which the slope of the Oslo method GSF will be further adjusted to. Moreover, the latter may require an additional scaling to match the absolute values of the Oslo method strength with the shape method result. The final assumption we make is, thus, that this scaling factor must remain within the limits set by the 95\% confidence interval of the $\langle \Gamma_{\gamma}\rangle$ value for $^{109}$In from the systematics. 

Calculations were performed for a wide range of $\sigma(S_n)$ values, from 3 to 7. For each value, the GSFs corresponding to transitions to the ground state ($D1$) and the cluster of the higher-lying states ($D2$) were extracted and scaled to the available photo-dissociation data, ($\gamma, xn$), for $^{115}$In at 9.5-10.5 MeV (see also Sec.\ref{sec 3: Results}). Further, the slope of the Oslo method GSF and the NLD was adjusted to the shape method result through a $\chi^2$ minimization, and the optimal BSFG parameters $a$ and $E_1$ reproducing the NLD with this slope were estimated. The scaling factor, adjusting the Oslo method strength to the shape method result, estimated for each $\sigma(S_n)$ is shown in Fig.~\ref{fig 3: Gg reduction factor}. The range of these factors corresponding to the 95\% confidence interval for $\langle \Gamma_{\gamma}\rangle$  extracted from the systematics is indicated in Fig.~\ref{fig 3: Gg reduction factor} by the horizontal shaded area. This range was used to determine the range of $\sigma(S_n)$ values and the recommended value of $\sigma(S_n)$ used to extract the final results (vertical shaded area in Fig.~\ref{fig 3: Gg reduction factor}). The recommended value corresponds to a reduction factor of 1.0, meaning that the Oslo and the shape method strengths align in absolute values after the slope adjustment, without any additional scaling. The recommended values of $\rho(S_n)$, $D_0$, $a$, and $E_1$  were also estimated for the recommended value of $\sigma(S_n)$. Their uncertainty ranges are directly determined by the span of $\sigma(S_n)$ values corresponding to the chosen $\langle \Gamma_{\gamma}\rangle$ scaling factor range (see Fig.~\ref{fig 3: Gg reduction factor}). All parameters determined in this procedure, as well as other parameters required for the normalization of the NLD are shown in Table \ref{tab:table_1}. 

\begin{figure}[t]
\includegraphics[width=1.0\linewidth]{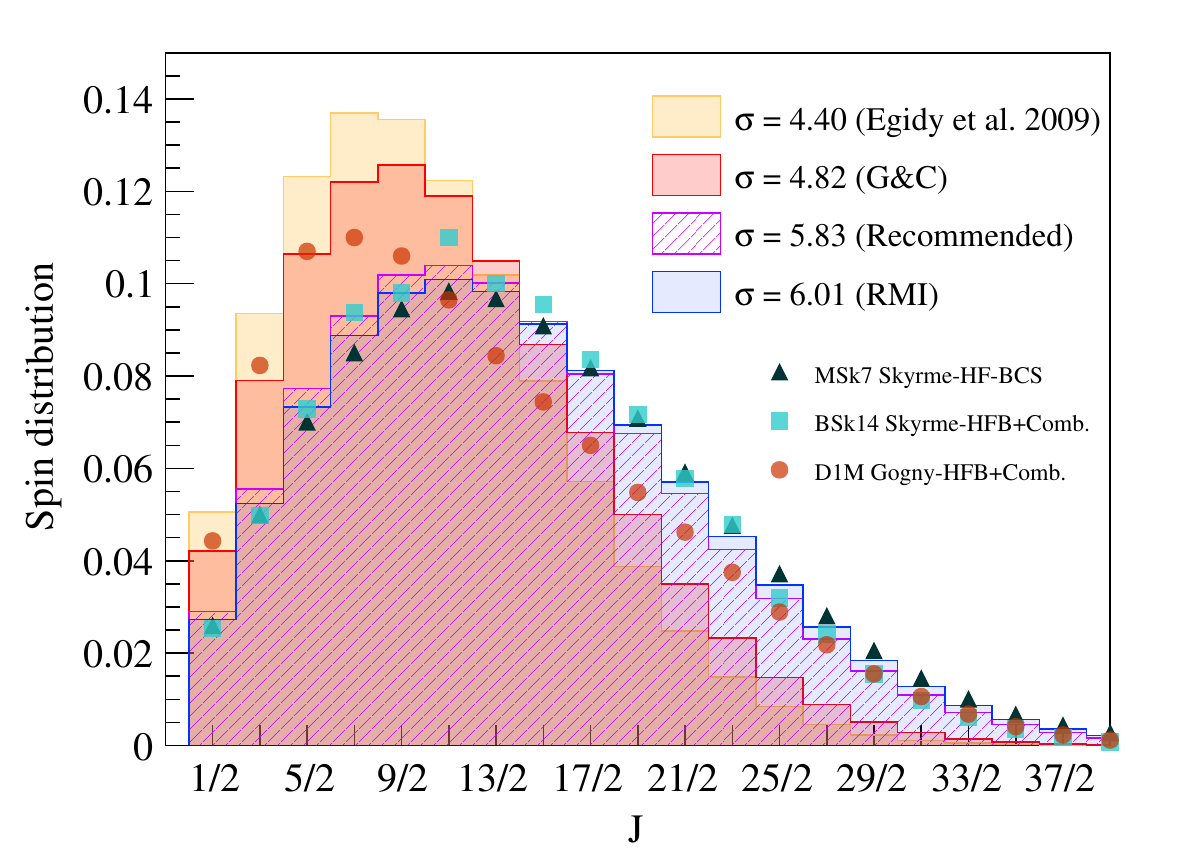}
\caption{\label{fig 4: Spin distribution} Spin distribution at the neutron separation energy estimated with the spin-cutoff provided by Eq.~(16) from Ref.~\cite{Egidy2009}, Eq.~(\ref{eq:8}) (G\&C) and Eq.~(\ref{eq:7}) (RMI) with the parameters $a$ and $E1$ taken from Ref.~\cite{Egidy05}. The spin-distribution based on the recommended $\sigma(S_n)$ value is presented by a hatched histogram. The spin-distributions from microscopic calculations with the Skyrme-HF-BCS plus statistical model based on the MSk7 Skyrme interaction \cite{Demetriou2001},  BSk14 Skyrme- and  D1M  Gogny-HFB plus combinatorial models (\cite{Goriely2008} and \cite{Hilaire2012}, respectively) are depicted with triangular, squared, and round markers.}
\end{figure}

\begin{table*}[t]
\caption{\label{tab:table_1}Parameters used for the normalization of the NLD and GSF of $^{109}$In. The values of $E_d$ and $\sigma_d$ were estimated from the low-lying discrete states in $^{109}$In \cite{ensdf}. Average total radiative width was estimated from the systematics for the even-odd Cd and Sn isotopes, the value is provided with its $2\sigma$ uncertainty.}
\begin{ruledtabular}
\begin{tabular}{lccccccccccc}
Nucleus & $S_n$ & $E_d$ & $\sigma_d$ & $\sigma(S_n)$ & $a$ & $E_1$ & $\rho(S_n)$ & $D_0$ & $\langle\Gamma_{\gamma}\rangle$ \\ 
& (MeV) & (MeV) & & & (MeV$^{-1}$) & (MeV) & ($ 10^6$ MeV $^{-1}$) & (eV) & (meV)\\
\noalign{\smallskip}\hline\noalign{\smallskip}
 $^{109}$In & 10.439 & 2.59(1) & 2.6(4) & 5.83(20) & 14.27$^{+0.15}_{-0.09}$ & $0.43^{+0.04}_{-0.03}$ & 7.2$^{+1.0}_{-0.6}$ & 1.45$^{+0.15}_{-0.19}$ &  111(15)\footnotemark[1] \\

\end{tabular}
\end{ruledtabular}
\footnotetext[1]{From systematics.}
\end{table*}
The estimated recommended value of $\sigma(S_n)$ overlaps well within its uncertainty with the RMI value and, therefore, results in a similar spin distribution at the neutron threshold (provided by Eq.~\ref{eq:5}), as shown in Fig.~\ref{fig 4: Spin distribution}. It is interesting to note that the spin distributions estimated with the microscopic Skyrme-Hartree-Fock (HF) plus
Bardeen-Cooper-Schrieffer (BCS) pairing plus statistical model \cite{Demetriou2001} and  Skyrme-Hartree-Fock-Bogoluybov (HFB) plus combinatorial \cite{Goriely2008} NLD calculations agree well with the spin distribution based on the recommended $\sigma(S_n)$ value. Other spin-cutoff values, commonly adopted for the Oslo method analysis in the earlier works, as well as Gogny-HFB plus combinatorial NLD calculations from Ref.~\cite{Hilaire2012} predict slightly lower average spins, cf. Fig.~\ref{fig 4: Spin distribution}.

\section{\label{sec 3: Results}Results and discussions}
\subsection{\label{subsec 3.1: NLD}Nuclear level density of $^{109}$In}

\begin{figure}[t]
\includegraphics[width=1.0\linewidth]{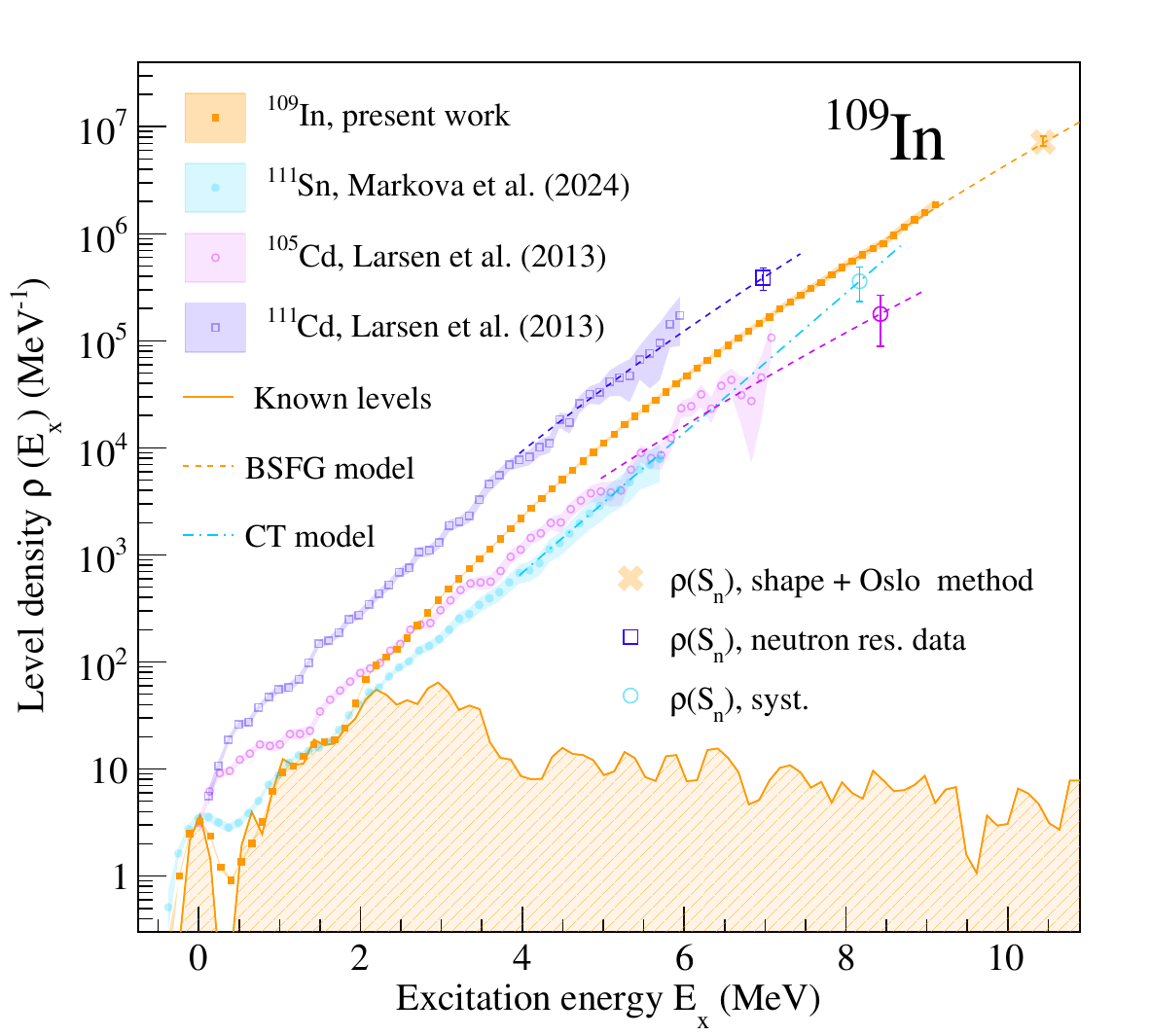}
\caption{\label{fig 5: NLD exp} Experimental NLD of $^{109}$In, plotted together with the NLDs of $^{105,111}$Cd \cite{Larsen2013_Cd} and $^{111}$Sn \cite{Markova2023, Markova2024}. Known low-lying states \cite{ensdf} are shown as a shaded hatched area. Dashed and dash-dotted lines indicate the BSFG and constant-temperature (CT) interpolations, respectively. Large open circles correspond to the $\rho(S_n)$ values extracted from the systematics of neutron resonance data. The open square and the cross correspond to $\rho(S_n)$ values from neutron resonance data and from the procedure described in Sec.\ref{subsec 2.3: Normalization 109In}, respectively. }
\end{figure}

\begin{figure}[t]
\includegraphics[width=1.0\linewidth]{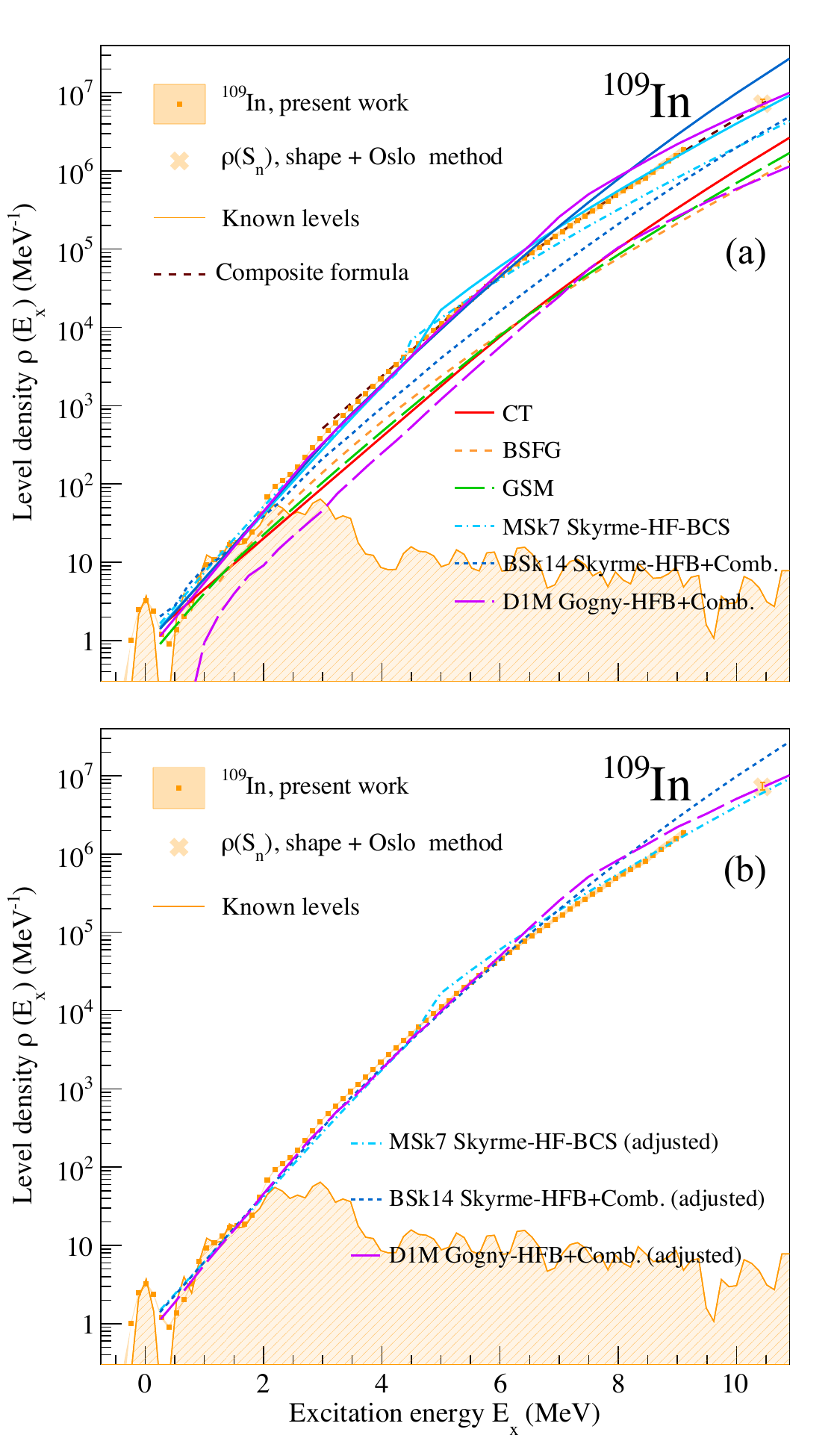}
\caption{\label{fig 6: NLD theor} (a) Experimental NLD of $^{109}$In plotted together with a fit with the composite formula \cite{Gilbert65}, CT, BSFG, Generalized Superfluid (GSM) phenomenological models, as well as calculations within the Skyrme-HF-BCS plus statistical \cite{Demetriou2001}, Skyrme-HFB plus combinatorial \cite{Goriely2008}, and Gogny-HFB plus combinatorial \cite{Hilaire2012} approaches (see Ref.~\cite{Koning2023} for more details regarding model calculations). No adjustments were made to the latter three model calculations. (b) Experimental NLD of $^{109}$In plotted together with the adjusted Skyrme-HF-BCS, Skyrme-HFB, and Gogny-HFB models (see text.)}
\end{figure}

The NLD of $^{109}$In extracted using the normalization procedure presented in the previous section is shown in Fig.~\ref{fig 5: NLD exp} together with the NLDs of the neighboring $^{105}$Cd, $^{111}$Cd, and $^{111}$Sn isotopes from the earlier works \cite{Larsen2013_Cd, Markova2023, Markova2024}. The statistical and systematic uncertainties were estimated in the same way as outlined in Refs.~\cite{Kullmann2019, Larsen2023, Markova2024}. The new results reproduce the low-lying discrete states up to approximately 2 MeV very well, where an abrupt, step-like increase of the level density is observed. At higher energies, the NLD follows closely the BSFG trend. To estimate the excitation energy range where the BSFG regime becomes applicable, the experimental results were fitted above 3 MeV with the so-called composite formula proposed in Ref.~\cite{Gilbert65}. This model assumes that the BSFG state is preceded by the CT regime at lower energies. The fit is indicated in Fig.~\ref{fig 6: NLD theor} by the dark-red dashed line. The BSFG appears to set in already at comparatively low energies of $\approx 5$ MeV. On the contrary, the chain of the Sn isotopes studied by the Oslo group \cite{Markova2024} displays a clear CT behavior up to $7-8$ MeV in some cases. For Cd isotopes \cite{Larsen2013_Cd}, the BSFG was used to interpolate the experimental data to the $\rho(S_n)$ values. It is, however, difficult to assess the transition point between the CT and BSFG regimes in these cases due to large error bars at high excitation energies.

The NLD of $^{109}$In appears to be significantly higher than that of the lightest studied Sn isotope, $^{111}$Sn. In general, the NLDs of even-odd Sn isotopes are close (within the uncertainties) in slopes and absolute values, primarily due to the stabilizing effect of the closed $Z=50$ shell. In terms of absolute values, much more variability is observed in Cd isotopes, also shown in Fig.~\ref{fig 5: NLD exp}. The newly extracted results for $^{109}$In lie well in between the NLDs of the closest neighboring even-odd $^{105}$Cd and $^{111}$Cd above approximately 4 MeV. In general, information on the NLDs of In isotopes is quite limited and consists of the level schemes at low excitation energies, as well as neutron resonance data for $^{113}$In and $^{115}$In targets, which can be converted to NLDs at the neutron separation energy in $^{114}$In and $^{116}$In, respectively. Information on the BSFG model parameters for $^{113,115}$In has been reported in Ref.~\cite{Lovtsikova1986}, suggesting even steeper slopes of the NLDs in these nuclei.

A comparison of the experimental NLD with phenomenological and microscopic models, commonly adopted for calculations with the reaction code \textsc{talys} (version 2.0) \cite{Koning2023}, is shown in Fig.~\ref{fig 6: NLD theor}(a). No additional adjustments were applied to the shown microscopic calculations. All of them, except for the Skyrme-HF-BCS plus statistical calculations \cite{Demetriou2001}, suggest much lower absolute values of the NLD below the neutron separation energy. The Skyrme-HF-BCS plus statistical model agrees quite well with the experiment up to approximately 6 MeV and deviates from it at higher energies. Analogous results are observed for the lightest studied $^{105}$Cd isotope (not shown in the figure). 

The three microscopic NLDs shown in Fig.~\ref{fig 6: NLD theor}(a) can be further adjusted to achieve better agreement with the experimental data. This was done by modifying the NLD slope and adding an energy shift according to Eq.(230) from Ref.~\cite{Koning2023}:
\begin{equation}
    \rho_{\rm mod}(E_x,J,\pi)=\exp(c\sqrt{E_x-\delta})\rho(E_x-\delta, J, \pi).
\end{equation}
Here, $\rho_{\rm mod}$ is the modified NLD, $c$ and $\delta$ are the modifying parameters. By requiring the best fit to the experimental data from 2.5 MeV to $S_n(^{109}$In)$=10.439$~MeV, the following coefficients were found: $c=0.33(5)$~MeV$^{-1/2}$, $\delta=0.37(4)$ MeV for the Skyrme-HF-BCS model, $c=0.64(4)$ MeV$^{-1/2}$, $\delta=0.36(4)$ MeV for the Skyrme-HFB model, $c=0.54(4)$ MeV$^{-1/2}$, $\delta=-0.54(2)$ MeV for the Gogny-HFB model. The adjustment of the latter model results in the best reproduction of the experimental data compared to all other considered theoretical approaches. In general, the modified Skyrme-HF-BCS and Gogny-HFB models reproduce the experimental data well from 2.5 MeV up to $S_n$ and support the $\rho(S_n)$ value found in the normalization procedure using the shape method. All adjusted calculations are shown Fig.~\ref{fig 6: NLD theor}(b) together with the experimental NLD.

In general, very little experimental information is currently available on NLDs of unstable nuclei, except for a few cases studied with the Oslo and the $\beta$-Oslo methods \cite{Spyrou2014}. In this respect, the $^{109}$In data presented in this work provide new valuable input for assessing the performance of different model approaches in the region of neutron-deficient heavy nuclei.

\subsection{\label{subsec 3.2: GSF}$\gamma$-ray strength function of $^{109}$In}

\begin{figure}[t]
\includegraphics[width=1.0\linewidth]{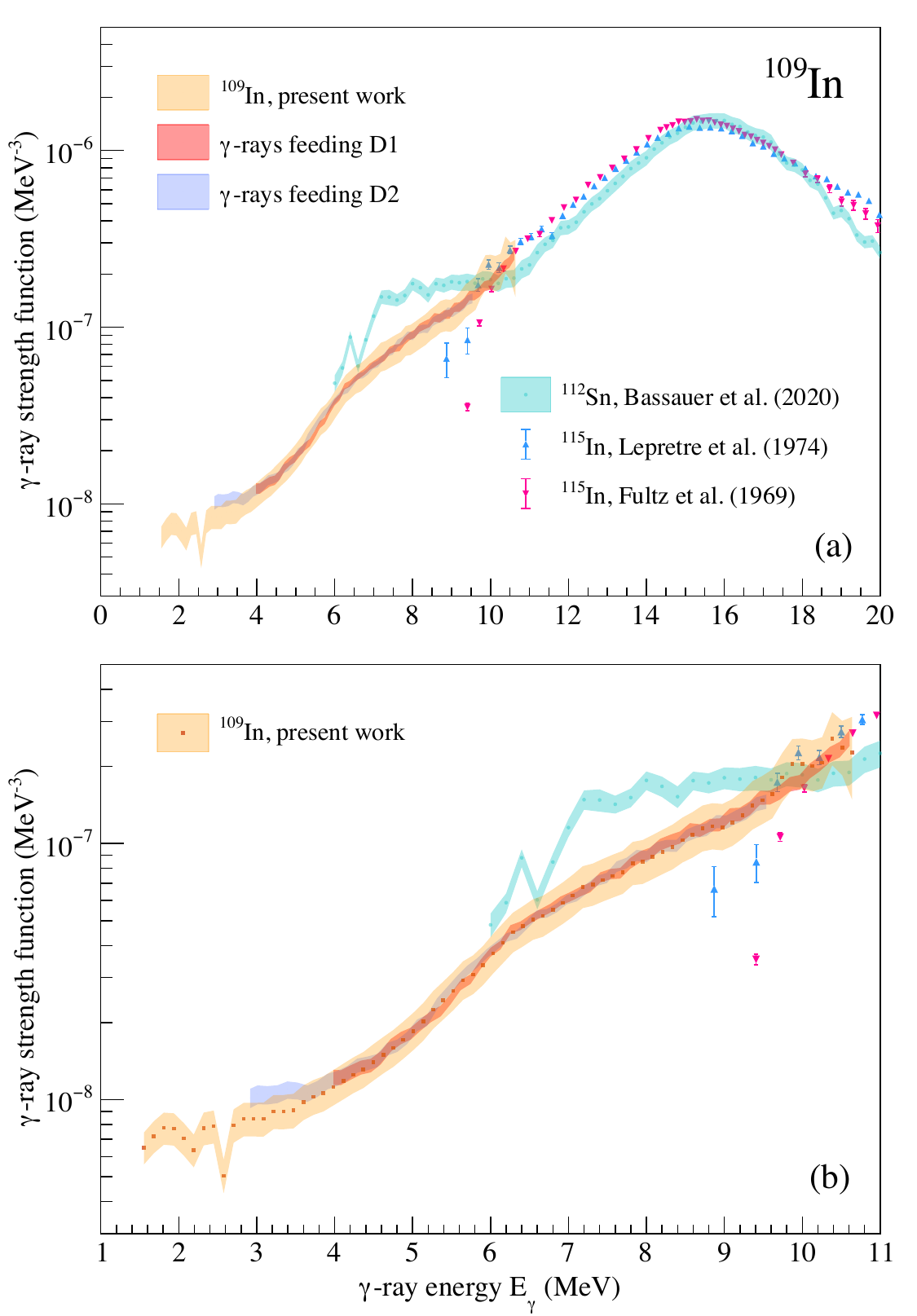}
\caption{\label{fig 7 1: GSF exp} (a) Oslo method GSF of $^{109}$In (orange band), shown together with the shape method results (red and blue shaded bands), the strength from the Coulomb excitation ($p,p^{\prime})$ experiment on $^{112}$Sn \cite{Bassauer2020b}, and photo-dissociation ($\gamma, xn$) data for $^{115}$In \cite{Lepretre74, Fultz69}. (b) Same as in (a) for $\gamma$-ray energies between 1~MeV and 10~MeV (the Oslo method data points are shown).}
\end{figure}

The GSF of $^{109}$In extracted with the Oslo method and the shape method is shown in Fig.~\ref{fig 7 1: GSF exp}. The GSF obtained with the shape method for $\gamma$-rays feeding the ground-state diagonal [D1 in Fig.~\ref{fig 1: Matrices}(c)] and the cluster of four low-lying states around 1.1 MeV [D2 in Fig.~\ref{fig 1: Matrices}(c)] using the recommended value of the spin-cutoff from Table~\ref{tab:table_1} is presented by red and blue bands in Fig.~\ref{fig 7 1: GSF exp}. The uncertainty of the recommended $\sigma(S_n)$ value was included in the shown uncertainty bands. The strengths overlap perfectly down to 4 MeV, chosen as the lower excitation-energy limit for the extraction of the GSFs. The shape method results cover a wide range of $\gamma$-ray energies (from 4 MeV to 10.4 MeV) and provide stringent constrains on the slope of the Oslo method GSF, thus yielding relatively small systematic uncertainties. The orange bands in Fig.~\ref{fig 7 1: GSF exp} correspond to the total uncertainties, including statistical uncertainties. After the slope adjustment, the overall shape of the Oslo method strength is in excellent agreement with the shape method GSF, in line with the adopted Brink-Axel hypothesis. A slight deviation, still within the uncertainty bands, of the GSF for the diagonal D2 from the Oslo method strength at approximately 3 MeV may potentially be attributed to transitions of $E2$ type, not accounted for in the shape method. 

\begin{figure}[t]
\includegraphics[width=1.0\linewidth]{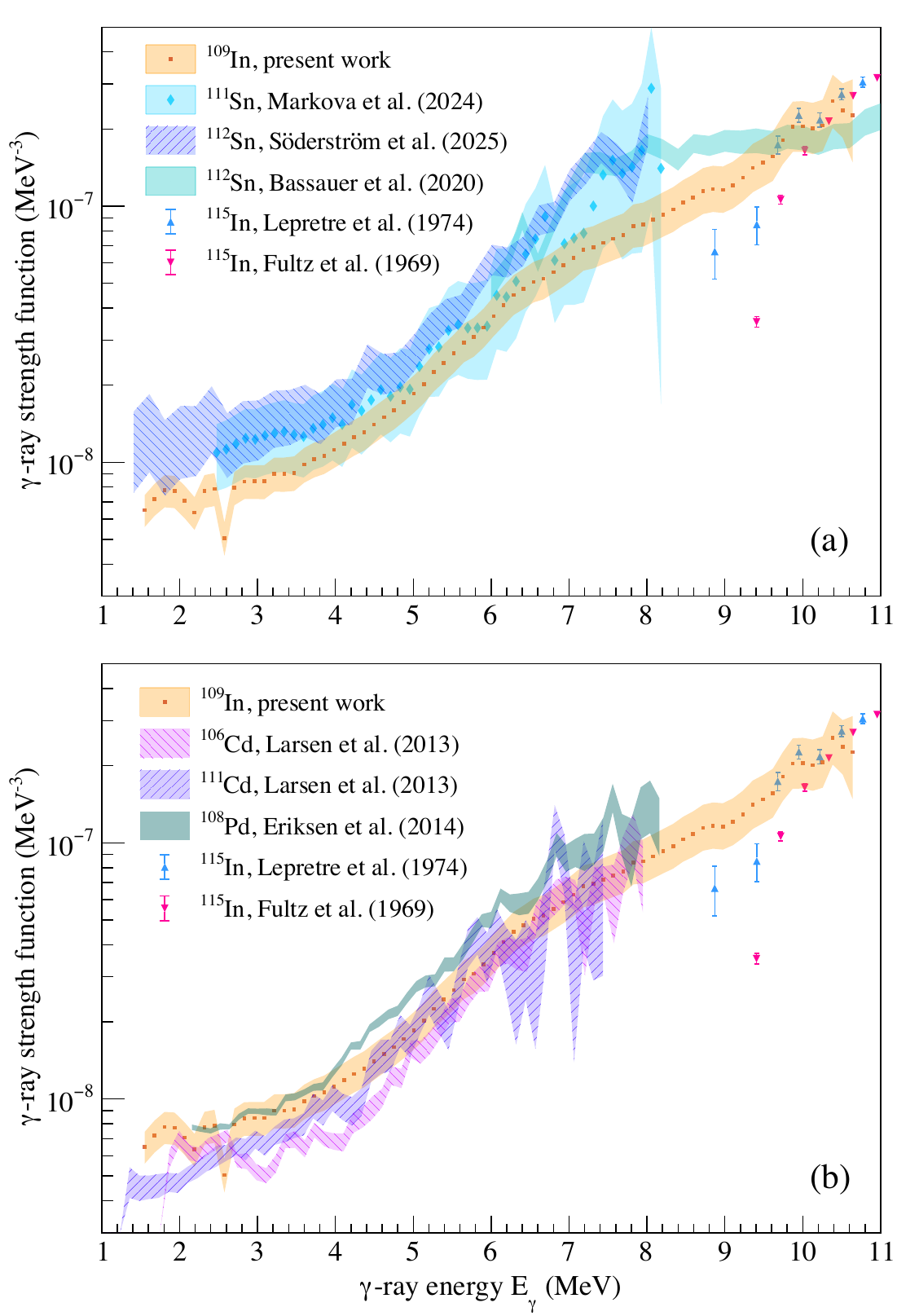}
\caption{\label{fig 7 2: GSF exp} (a) Comparison of the Oslo method GSF of $^{109}$In (orange band with squared markers) with the Oslo method strengths of the neighboring Sn isotopes, $^{111}$Sn and $^{112}$Sn from the ($p,d\gamma$) \cite{Markova2023, Markova2024} and ($p,p^{\prime}\gamma$)  \cite{Soderstrom2025} experiments, respectively, shown together with the Coulomb excitation data for $^{112}$Sn \cite{Bassauer2020b} and photo-dissociation ($\gamma, xn$) data for $^{115}$In \cite{Lepretre74, Fultz69}. (b) Comparison of the Oslo method $^{109}$In result with the $^{106,111}$Cd \cite{Larsen2013_Cd} and $^{108}$Pd \cite{Eriksen2014} Oslo method strengths from the $^3$He-induced reactions, as well as the ($\gamma, xn$) data for $^{115}$In.}
\end{figure}

\begin{figure}[t]
\includegraphics[width=1.0\linewidth]{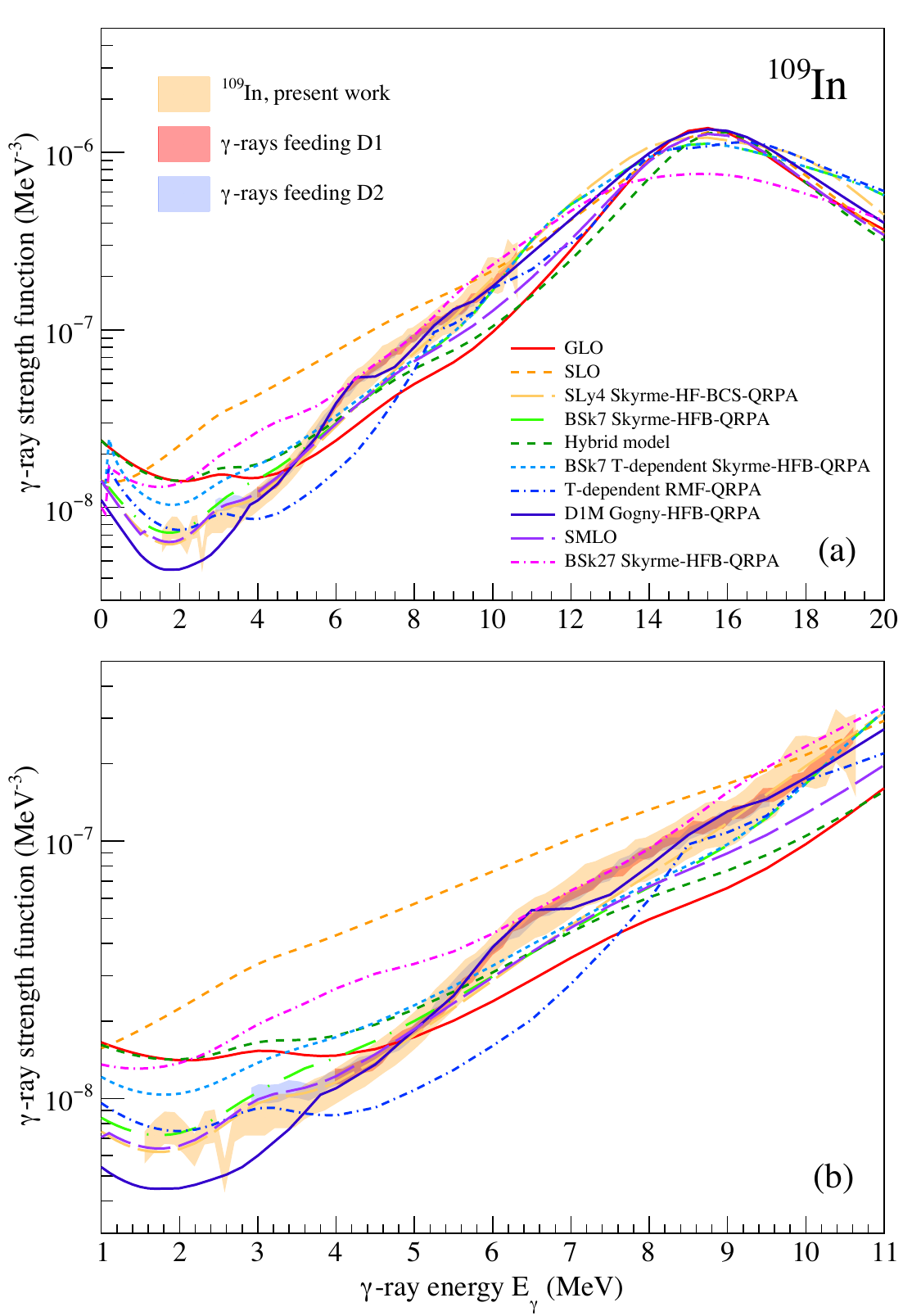}
\caption{\label{fig 8: GSF theor} Experimental GSF of $^{109}$In extracted with the Oslo and shape methods, shown together with the dipole strength model predictions with the $E1$ component based on the generalized (GLO) and simple (SLO) Lorentzians \cite{Koning2023}, the hybrid model \cite{Goriely98}, Skyrme-HF-BCS-QRPA \cite{Goriely2002}, Skyrme-HFB-QRPA \cite{Goriely2004, Xu2021}, temperature-dependent Skyrme-HFB-QRPA \cite{Goriely2004}, temperature-dependent relativistic mean field (RMF)-QRPA \cite{Daoutidis2012}, Gogny-HFB-QRPA \cite{Goriely2018}, Simple Modified Lorentzian (SMLO) \cite{Goriely2019}. (b) Same as in (a) for $\gamma$-ray energies between 1 MeV and 10 MeV.}
\end{figure}

As mentioned earlier, the shape method results were scaled to the $(\gamma, nx)$ data for $^{115}$In \cite{Fultz69, Lepretre74} at approximately 10 MeV. At these energies, the low-energy flank of the IVGDR  appears to be quite similar across the chain of the neighboring Sn isotopes studied in Coulomb excitation experiments \cite{Bassauer2020b}. We therefore assume that the same holds for the In isotopes, and that the low-energy part of the IVGDR at 10 MeV varies little between $^{109}$In and $^{115}$In.

The comparison with other experimental data, in particular the earlier Oslo method results, is shown in Fig.~\ref{fig 7 2: GSF exp}. Here, the shape method strengths for $^{109}$In are omitted as they are in excellent agreement with the Oslo method data. The $^{109}$In and the $^{111}$Sn \cite{Markova2023, Markova2024} strengths are in fairly good agreement from approximately 4 to 7 MeV within the uncertainty bands. The recent result by Söderström \textit{et al.} \cite{Soderstrom2025} exhibits a similar shape but a somewhat higher strength across the whole shown energy range. An overall good agreement above 5 MeV is also observed between $^{109}$In and the closest earlier studied $^{106,111}$Cd isotopes \cite{Larsen2013_Cd}. The Pd data from the study by Eriksen \textit{et al.} \cite{Eriksen2014} are also shown in Fig.~\ref{fig 7 2: GSF exp}(b). The shapes of GSFs of the $^{105-108}$Pd isotopes from this work are quite similar, and, therefore, only the $^{108}$Pd case is shown here. These data agree quite well with $^{109}$In at low energies but, similar to the new $^{112}$Sn result [see Fig.~\ref{fig 7 2: GSF exp}(a)], demonstrate a slightly larger strength at higher energies.

This generally good agreement with the strengths of the neighboring nuclei indirectly supports the normalization of the $^{109}$In data, in particular the resulting slopes of the NLD and the GSF.  At lower energies, while still agreeing with the $^{111}$Sn data within their rather broad uncertainty band, $^{109}$In falls well in between the Cd and Sn cases, exhibiting no signs of a low-energy enhancement, the so-called upbend, observed in the lighter Cd isotopes \cite{Larsen2013_Cd}. 

An interesting feature observed in the new $^{109}$In data is the lack of any apparent enhancement in the vicinity of 8 MeV, which is clearly seen in the Sn isotopes (see Fig.~\ref{fig 7 1: GSF exp}, Fig.~\ref{fig 7 2: GSF exp}(a), and Refs.~\cite{Bassauer2020b, Markova2024}) and hinted at in the Cd and Pd data (see  Fig.~\ref{fig 7 2: GSF exp}(b) and Refs.~\cite{Larsen2013_Cd, Eriksen2014}). This feature is essentially independent of the normalization procedure and makes the GSF of $^{109}$In stand out, in terms of shape, from the neighboring Pd, Cd, Sn, and Sb isotopes. Due to the absence of this 8-MeV enhancement, associated with the PDR in the earlier works ~\cite{Larsen2013_Cd, Eriksen2014}, some of the theoretical model calculations employed in \textsc{talys} are much more compatible with the experimental data for $^{109}$In than in other cases in this mass region (see, e.~g., \cite{Pogliano2022}). The comparison of phenomenological models and calculations within the quasi-particle random-phase approximation (QRPA) with the experiment is shown in Fig.~\ref{fig 8: GSF theor}. Here, all options listed in the figure legend, except for the Gogny-HFB-QRPA \cite{Goriely2018} and the most recent Skyrme-HFB-QRPA \cite{Xu2021}, were combined with the default \textsc{talys} $M1$ model (SMLO from Ref.~\cite{Goriely2019}) to produce the total dipole strength for the comparison with the experimental data. The Gogny-HFB-QRPA and Skyrme-HFB-QRPA $E1$ strengths were combined with the corresponding $M1$ strengths obtained within the same frameworks. In all cases, an upbend prescribed in Ref.~\cite{Koning2023} was added to the shown calculations. The Gogny-HFB-QRPA and Skyrme-HF-BCS-QRPA approaches agree nicely with the Oslo data below the neutron threshold within the experimental uncertainties. Substantial deviation of other theoretical strengths from each other and the experimental results is still observed, highlighting the importance of our new data in constraining theoretical uncertainties related to the GSF models, in much the same way as for the NLDs.

\begin{table*}[t]
\caption{\label{tab:table_2} Parameters used for the description of the IVGDR (Generalized Lorentzian), M1 strength (Lorentzian), low-lying $E1$ strength (Gaussian), and the upbend (exponential function) in $^{109}$In. The notations are kept the same as in Ref.~\cite{Markova2024}.}
\begin{ruledtabular}
\begin{tabular}{cccccccccccc}
\multicolumn{4}{c}{IVGDR} & \multicolumn{3}{c}{$M1$} & \multicolumn{2}{c}{Upbend} & \multicolumn{3}{c}{low-energy $E1$}\\\noalign{\smallskip}
\cline{1-4} \cline{5-7} \cline{8-9} \cline{10-12}  \noalign{\smallskip}
$E_{E1}$ & $\Gamma_{E1}$ & $\sigma_{E1}$ & $T_{f}$ & $E_{M1}$ & $\Gamma_{M1}$ & $\sigma_{M1}$ & $C_{\rm up}$ & $\eta_{\rm up}$ & $E_{E1}^{\rm low}$ & $\sigma_{E1}^{\rm low}$ & $C_{E1}^{\rm low}$ \\ 
(MeV) & (MeV) & (mb) & (MeV) & (MeV) & (MeV) & (mb) & (10$^{-8}$ MeV$^{-3}$) & (MeV$^{-1}$) & (MeV) & (MeV) & ($10^{-7}$ MeV$^{-2}$)\\
\noalign{\smallskip}\hline\noalign{\smallskip}
16.22(2) & 7.0(1) & 238.6(14) & 0.40(3) & 11.5(3) & 5.6(5) & 1.6(2) & 1.4(7) & 1.2(4) & 7.9(4) & 1.4(3) & 1.0(2)\\
\end{tabular}
\end{ruledtabular}
\end{table*}

Despite no clear 8-MeV enhancement in the strength, it might still be instructive to perform a decomposition of the total dipole strength, provided by the Oslo and ($\gamma, xn$) data in terms of the IVGDR, the low-lying $E1$ strength, and the $M1$ component in a similar way as was done in Refs.~\cite{Larsen2013_Cd, Eriksen2014, Pogliano2022, Markova2024}. Here, we are employing the same functions to parametrize these features as in the previous works: the generalized Lorentzian function, characterized by centroid energy $E_{E1}$, cross-section $\sigma_{E1}$, width $\Gamma_{E1}$, and temperature $T_{E1}$ for the IVGDR, the standard Lorentzian function with energy centroid $E_{M1}$, cross-section $\sigma_{M1}$, and width $\Gamma_{M1}$ for the $M1$ response, and the Gaussian function with energy centroid $E_{E1}^{low}$, scaling factor $C_{E1}^{low}$, and width $\sigma_{E1}^{low}$ for the low-lying $E1$ enhancement on top of the IVGDR. For consistency, we use the same formulae and notations as in the most recent work \cite{Markova2024} [Eq.~(13)-(16)]. The $^{115}$In data are used to represent the IVGDR in $^{109}$In. Since experimental information on the $M1$ strength in In nuclei is lacking, we adopt the same systematics for the $M1$ response from Coulomb excitation experiments \cite{Bassauer2020b} as those used in Ref.~\cite{Markova2024}. For a better reproduction of the experimental data at low $\gamma$-ray energies, we add an additional function, $C_{\rm up}\exp(-\eta_{\rm up}E_{\gamma})$, characterized by a scaling $C_{up}$ and a slope $\eta_{up}$ parameter. This function is commonly used to model the low-energy enhancement in cases where it is observed (see, e.g.,~\cite{Guttormsen2022}). In $^{109}$In, similarly to the Sn isotopes, it has no significant effect on the fit results.

\begin{figure}[b]
\includegraphics[width=1.0\linewidth]{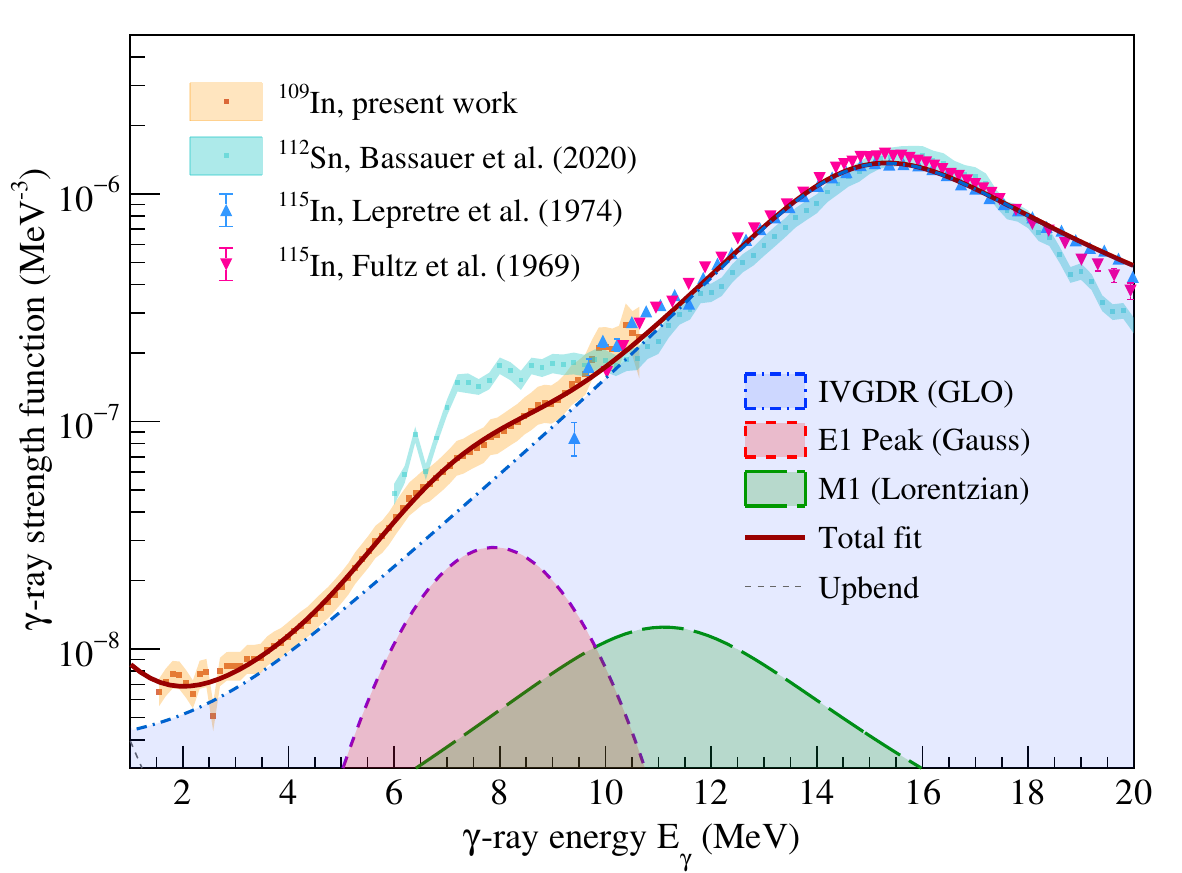}
\caption{\label{fig 9: GSF decomposition} Decomposition of the dipole strength into the IVGDR, the $M1$ response, and the low-lying $E1$ strength on top of the tail of the IVGDR. The total fit is shown as a solid dark-red line. The Coulomb excitation data for $^{112}$Sn are included for comparison.}
\end{figure}

\begin{figure}[h]
\includegraphics[width=1.0\linewidth]{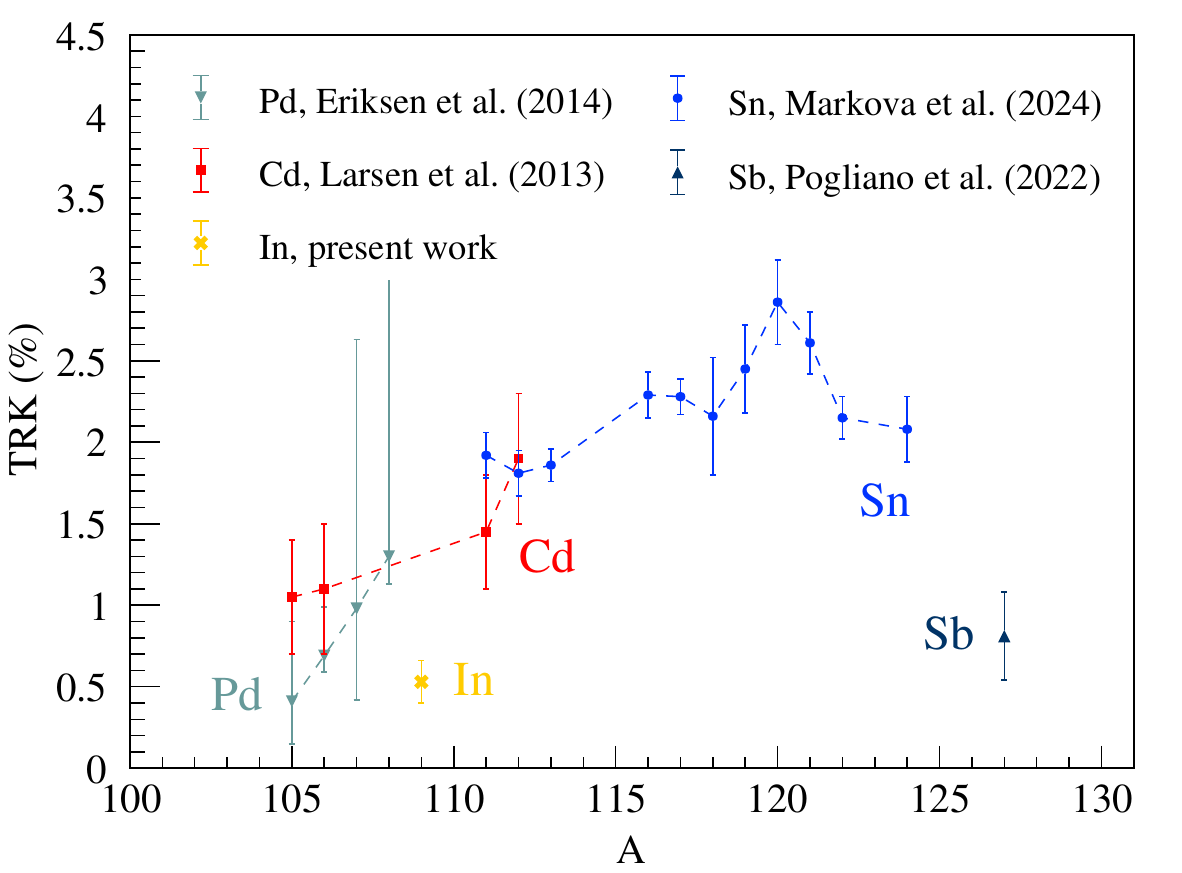}
\caption{\label{fig 10: LEDR systematics} TRK values exhausted by the low-lying $E1$ strength in $^{105-108}$Pd \cite{Eriksen2014}, $^{105,106,111,112}$Cd \cite{Larsen2013_Cd}, $^{111-113,116-122,124}$Sn \cite{Markova2024}, $^{127}$Sb (average of two reported values in Ref.~\cite{Pogliano2022}), and $^{109}$In. Dashed lines are added as a visual aid.}
\end{figure}

\begin{figure*}[t]
\includegraphics[width=0.9\linewidth]{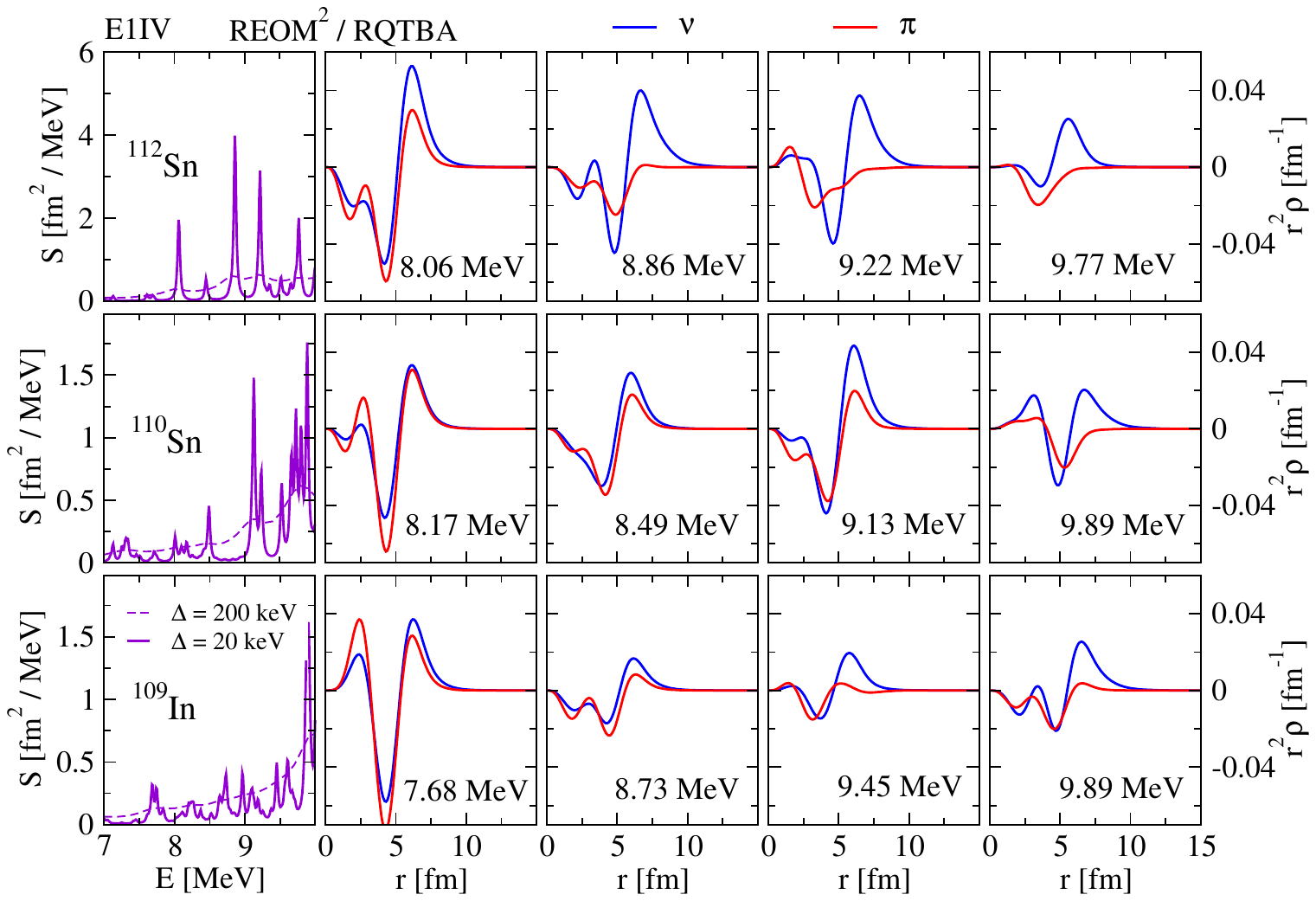}
\caption{REOM$^2$/RQTBA E1 strength functions below 10 MeV in $^{112}$Sn, $^{110}$Sn and $^{109}$In (left panels), and the neutron and proton transition densities for the strongest four peaks of the corresponding strength function (four right panels in each raw) with the indicated energies.}
\label{TrDen}
\end{figure*}

The Oslo and the $(\gamma, xn)$ data were fitted simultaneously by a composite function consisting of all the above-mentioned terms. The total fit and the decomposition of the experimental strength in terms of its main components are shown in Fig.~\ref{fig 9: GSF decomposition}, and the fit parameters with the uncertainties are provided in Table \ref{tab:table_2}. The low-lying $E1$ strength extracted in this way is considerably smaller than in the neighboring Pd, Cd, and Sn nuclei and exhausts only 0.53(13)\% of the Thomas-Reiche-Kuhn (TRK) sum rule \cite{Thomas1925,Reiche1925,Kuhn1925}. The TRK values corresponding to the additional strength on the top of the IVGDR extrapolation to low energies in Pd, Cd, In, Sn, and Sb nuclei are shown in Fig.~\ref{fig 10: LEDR systematics} as a function of mass number $A$. As was discussed in Refs.~\cite{Markova2024, Markova2025}, this strength should not be considered representative of the PDR mode in its neutron-skin interpretation, and one should therefore not expect any correlation with neutron excess. This is supported by the Sn data, where the integrated low-lying strength remains relatively constant from $^{111}$Sn to $^{124}$Sn. In Cd and Pd isotopes, the increase of the strength reported in Ref.~\cite{Eriksen2014} remains speculative due to the lack of data around 8 MeV and the corresponding large uncertainties. The low-lying $E1$ strengths in the neighboring Pd, Cd, and Sn isotopes are rather similar, in both the shapes of their GSFs and the extracted TRK values (cf. $^{111,112}$Cd and $^{111,112}$Sn in Fig.~\ref{fig 10: LEDR systematics}). Although, one might expect the strength in In isotopes to be compatible with that of immediate neighbors from other isotopic chains, the new $^{109}$In data demonstrate that this is not the case. It is important to emphasize that the TRK value for $^{109}$In strongly depends on the model for the $M1$ response adopted in the decomposition of the total dipole strength. Any of $M1$ resonance parametrizations suggested, e.~g., in Ref.~\cite{Koning2023} would yield even lower TRK fractions for $^{109}$In. The value reported in this work should be therefore regarded as an attempt to assess the low-lying $E1$ strength in $^{109}$In on the same footing as for the Sn and Sb isotopes and to highlight the reduced strength in the present case compared to its neighbors. It is interesting to note that the TRK value extracted for $^{127}$Sb is also comparatively low, despite its GSF being quite similar to that of $^{124}$Sn. This can be attributed to the modeling of the IVGDR and the fitting procedure adopted in Ref.~\cite{Pogliano2022}.

\subsection{\label{subsec 3.3: RQTBA} Theoretical model calculations}

Calculations beyond QRPA were performed within the relativistic equation of motion (REOM)
approach, which represents a consolidation of the long-term effort on the nuclear many-body theory and its numerical implementations \cite{PLANT1998607,Ring1996,VretenarAfanasjevLalazissisEtAl2005,PaarRingNiksicEtAl2003,LitvinovaRingTselyaev2008,Tselyaev2013,LitvinovaSchuck2019,Litvinova2022}. 
Already in the leading approximation to the quasiparticle-vibration coupling (qPVC) dubbed as  REOM$^2$/RQTBA, considerable improvements of the GSFs were achieved compared to QRPA \cite{Markova2024,Markova2025}.
As in Ref. \cite{Markova2025}, here the REOM$^2$/RQTBA, based on exact ab-initio many-body theory and underlying meson-exchange interaction, adopts an implementation where the nuclear excited-state wave functions are approximated by two-quasiparticle-plus-phonon ($2q\otimes phonon$) configurations. The latter causes fragmentation of the QRPA states composed of pure $2q$ configurations, generating the spreading width of the IVGDR and the PDR strength distribution.

Here we perform a comparative analysis of the low-energy E1 strength below 10 MeV computed for $^{112}$Sn, $^{110}$Sn and $^{109}$In. While the relativistic QRPA (RQRPA) yields a single peak in each of these nuclei at 9.1 MeV, 9.3 MeV, and 9.4 MeV, respectively, in this energy region, the inclusion of qPVC produces multitudes of states fragmented from these ``primordial" RQRPA modes. The REOM$^2$/RQTBA strength distributions are collected in the left panels of Fig.~\ref{TrDen}.  The first observation is that the total strength below 10 MeV drops significantly when going from $^{112}$Sn to $^{110}$Sn (note the different scale for $^{112}$Sn) and further decreases in $^{109}$In. Remarkably, the shell structure of the Dirac-Hartree basis states is very similar in all three nuclei: the proton subsystem is characterized by the closed 1g$_{9/2}$ shell with one proton removed from this orbit in $^{109}$In, and the neutron subsystem has filled 1g$_{7/2}$ (eight particles) above the $N = 50$ core in all the nuclei and a partly occupied 2d$_{5/2}$ orbit with two neutrons in $^{110}$Sn and $^{109}$In and four neutrons in $^{112}$Sn. However, this relatively subtle difference in the occupancy results in a notable difference in the transition strength of the lowest RQRPA mode, much larger in 
$^{112}$Sn due to the predominant contribution of the 2d$_{5/2}\to$2f$_{7/2}$ neutron transition.

This enhancement is translated to the REOM$^2$/RQTBA E1 strength functions, whose composition can be traced via the transition densities. We display them in Fig.~\ref{TrDen} for the four most prominent fragments seen in the low-energy dipole strength function of each nucleus. At lower energies, proton and neutron contributions closely follow each other, and here we remind that the transition strength in the isovector channel is obtained by summing up these transition densities with nearly equal-valued opposite charges and integration over the radial distance with the dipole operator. So, the states below 9 MeV in $^{110}$Sn and $^{109}$In and below 8.5 MeV in $^{112}$Sn are dominated by destructive interference. Above these energies,
the increasing role of the neutron contribution when going from $^{109}$In to $^{112}$Sn is clearly visible. Further, one can notice that the oscillating behavior of the individual transition densities also cancels them considerably when the integration is performed.

%

\begin{figure}
\includegraphics[width=1.0\linewidth]{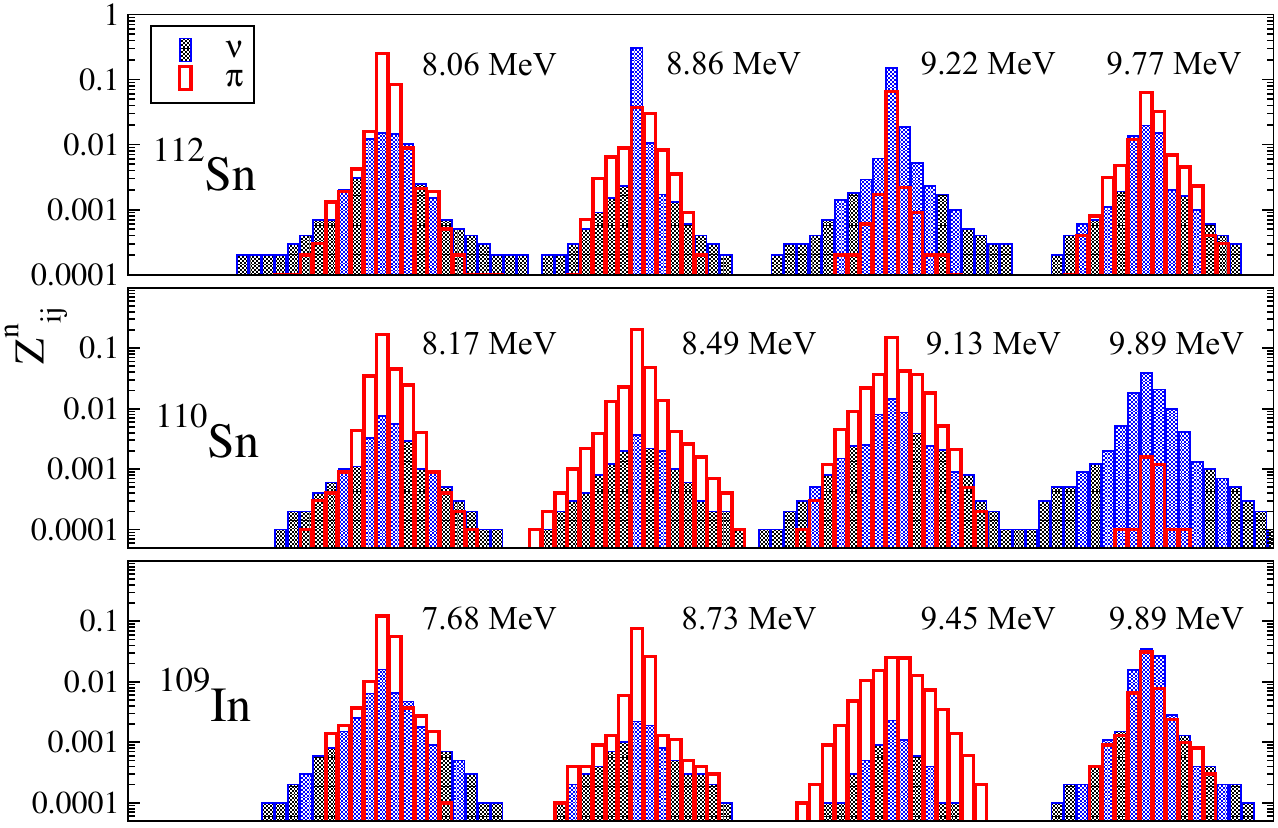}
\caption{Contributions of the leading (larger than 0.01\%) $2q$ configurations to the norms of the characteristic states in $^{112}$Sn, $^{110}$Sn and $^{109}$In below 10 MeV.}
\label{Ampls}
\end{figure}

Therefore, Fig.~\ref{Ampls} provides a complementary view of the composition of these states. Here, we plotted the quantity $Z^n_{ij} = |{\cal X}^n_{ij}|^2 - |{\cal Y}^n_{ij}|^2$, which characterizes the contribution of the $2q$ configuration formed by two fermions on orbits $|i\rangle$ and $|j\rangle$ to the norm of the many-body excited state $|n\rangle$. Note that in REOM$^2$/RQTBA
the summed $Z^n$-values over the $2q$ pairs $\{ij\}$ exhaust only a part of the state norm, while the remaining part includes the $2q\otimes phonon$ configurations \cite{LitvinovaRingTselyaev2007,Litvinova2007}. However, only the $2q$ portion couples to the leading one-body excitation operator, which explains the quenched strength of the fragmented states, so that this portion contributes only to one-body transition densities and amplitudes collected here.  Remarkably, as seen in Fig.~\ref{Ampls}, all the characteristic states in $^{109}$In are dominated by proton contributions, except for the one at 9.89 MeV where the leading proton and neutron $2q$ amplitudes are comparable. A similar situation is observed in $^{110}$Sn, where a stronger neutron contribution appears only at the end of the interval. In contrast, in $^{112}$Sn only the lowest and highest states at 8.06 MeV and 9.77 MeV are proton-dominated, while the most prominent peaks located at 8.86 MeV and 9.22 MeV exhibit a sizeable neutron dominance which can be attributed to the neutron oscillation at the surface visible in Fig. \ref{TrDen}. 

Thus, within the REOM$^2$/RQTBA model, we obtain a qualitative change of the pattern of the low-energy IV $E1$ strength. In contrast to neutron-rich nuclei, in the neutron-deficient species, the lowest dipole states are largely driven by proton transitions. Although the neutron peak at the surface, often associated with the neutron skin oscillation, is still notably expressed in the radial behavior of the transition densities, its surface contribution to the strength is compensated by the oscillation in the interior. The $^{110}$Sn nucleus thus represents the onset of this qualitative change within the tin isotopic chain.

\section{\label{sec 4: Astro}Neutron- and proton-capture cross sections and rates}

\begin{figure}[t]
\includegraphics[width=1.0\linewidth]{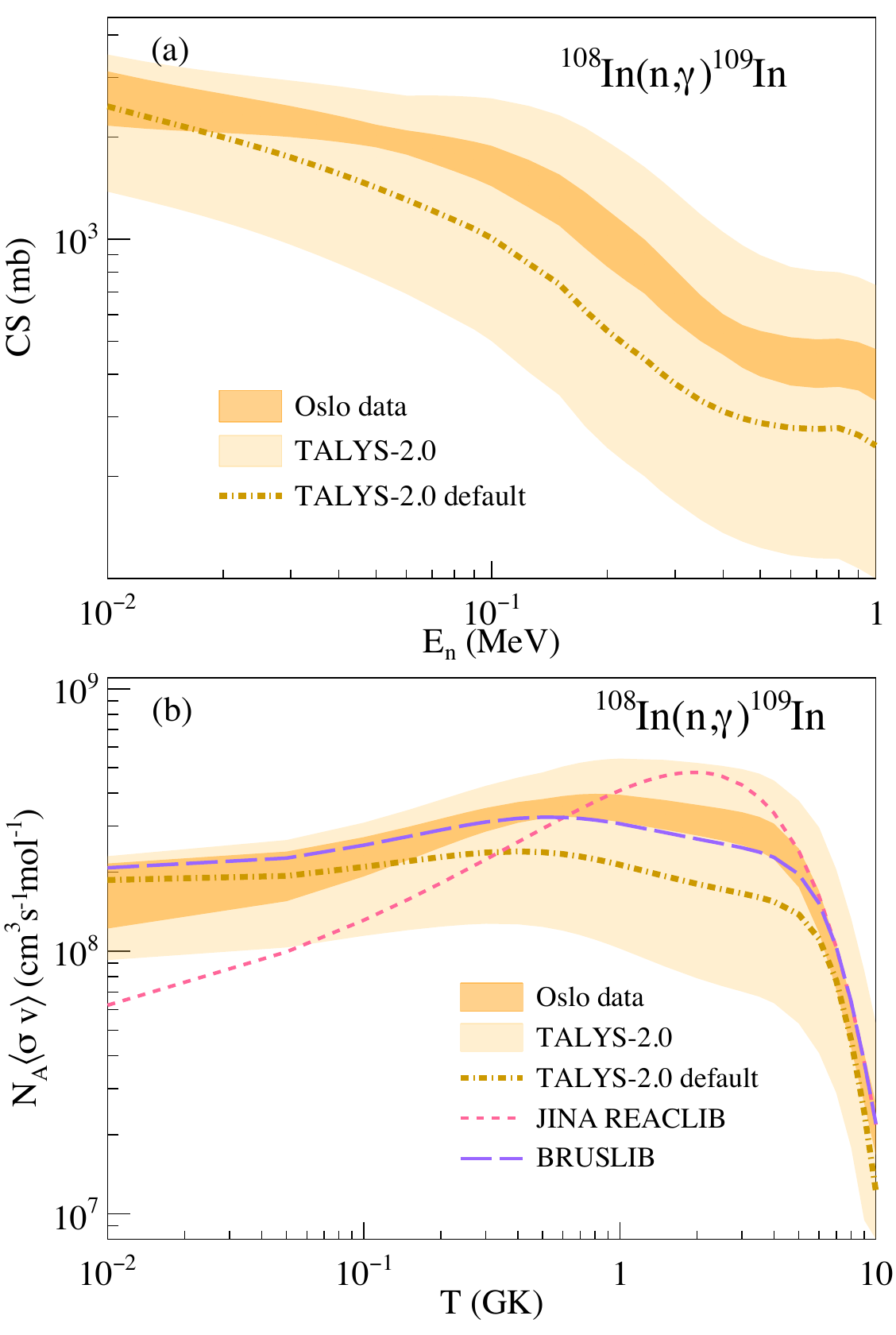}
\caption{\label{fig 10: ng CS and Rates}
The experimentally constrained cross section (a) and rate (b) of the $^{108}$In($n, \gamma$)$^{109}$In reaction, (orange band) shown together with the \textsc{talys} model uncertainty (beige band) and default calculations (dash-dotted line). Predictions from the JINA REACLIB \cite{JINA} and BRUSLIB \cite{BRUSLIB} libraries are displayed in the bottom panel with short-dashed and long-dashed lines, respectively.
}
\end{figure}

\begin{figure}[t]
\includegraphics[width=1.0\linewidth]{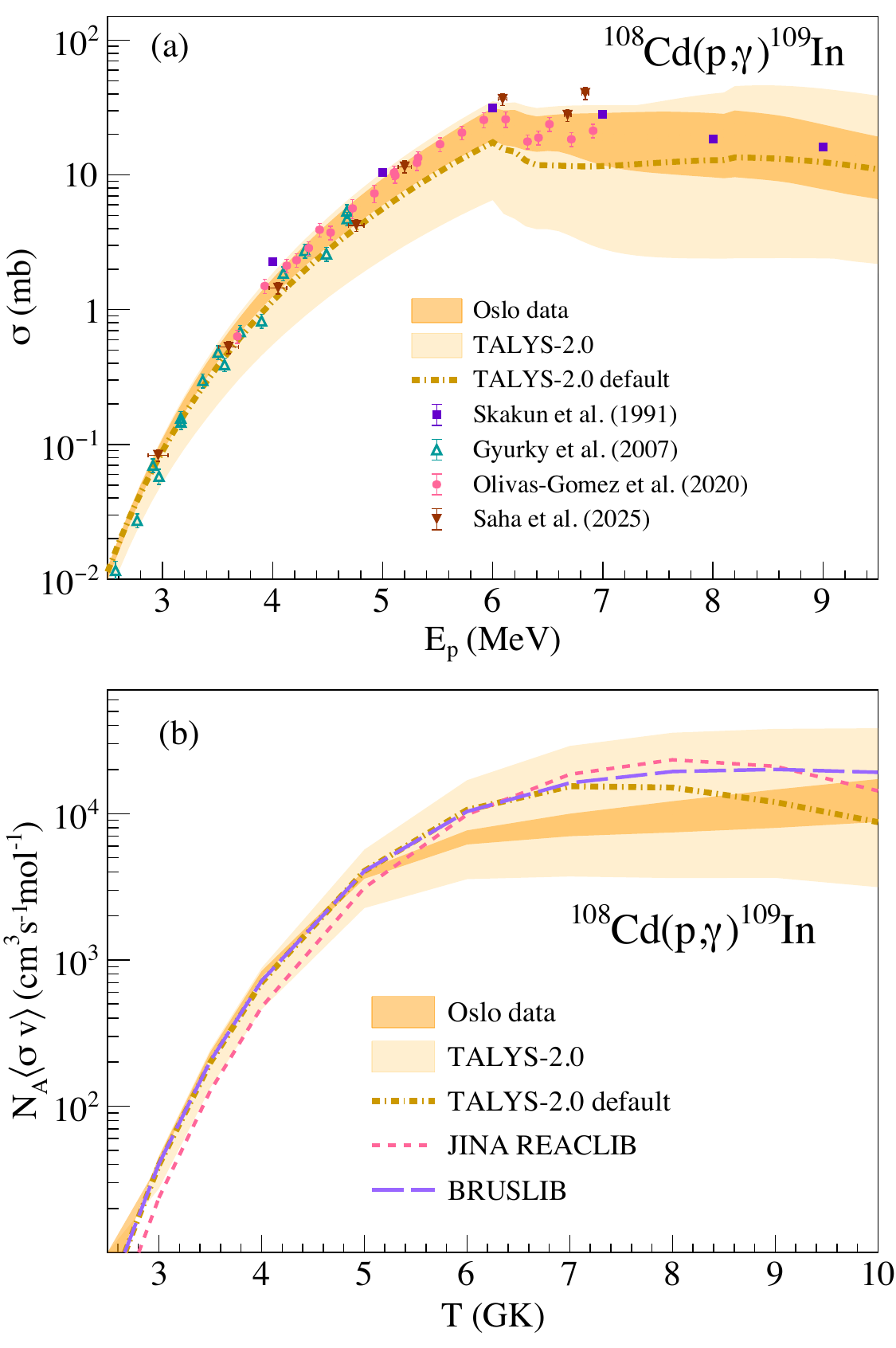}
\caption{\label{fig 11: pg CS and Rates}
The experimentally constrained cross section (a) and rate (b) of the $^{108}$Cd($p, \gamma$)$^{109}$In reaction, (orange band) shown together with the \textsc{talys} model uncertainty (beige band) and default calculations (dash-dotted line). The experimental data by Skakun \textit{et al.} \cite{Skakun1992}, Gyürky \textit{et al.} \cite{Gyürky2007}, Olivas-Gomez \textit{et al.} \cite{Olivas-Gomez2020}, and  Saha \textit{et al.} \cite{Saha2025} are shown in the upper panel. Predictions from the JINA REACLIB \cite{JINA} and BRUSLIB \cite{BRUSLIB} libraries are displayed in the bottom panel with short-dashed and long-dashed lines, respectively.
}
\end{figure}

As was mentioned in the introduction, the NLDs and GSFs extracted with the Oslo method can provide important constraints for calculations of radiative neutron- and proton-capture rates in astrophysical $p$, $s$, $i$, and $r$ processes within the Hauser-Feschbach statistical framework \cite{Larsen2019}. To address the impact of our new experimental data for $^{109}$In, we performed calculations of cross sections and rates of the radiative neutron- and proton-capture reactions, $^{108}$In($n, \gamma$)$^{109}$In and $^{108}$Cd($p, \gamma$)$^{109}$In, respectively, using the \textsc{talys} software. The experimental Oslo method NLD and GSF, combined with the ($\gamma, xn$) data for $^{115}$In at energies above the neutron separation energy, were used to produce tabulated NLD and $E1$ strength inputs for \textsc{talys} calculations. In addition to the uncertainties of all normalization parameters, the uncertainty in the $M1$ response and, therefore, the decomposition of the experimental strength into the $E1$ and $M1$ components was taken into account: all $M1$ models available in \textsc{talys}, as well as the $M1$ resonance extracted from the Sn systematics (see Fig.~\ref{fig 9: GSF decomposition}), were included in the uncertainty band. Moreover, the optical potentials from both the phenomenological model of Koning and Delaroche \cite{Koning2003} and the semi-microscopic Jeukenne-Lejeune-Mahaux model renormalized by the Bruy\`eres-le-Ch\^atel group \cite{Bauge2001} were included in the experimental uncertainties for the cross sections. The results of the calculations with the experimental Oslo method inputs are shown as shaded yellow bands in Figs.~\ref{fig 10: ng CS and Rates} and \ref{fig 11: pg CS and Rates} for neutron and proton capture reactions, respectively. The \textsc{talys} model uncertainty in both cases was also estimated by varying all available NLD, optical model potential, $E1$ and $M1$ strength options. When considering different model combinations, the Skyrme-HFB-QRPA \cite{Xu2021} and Gogny-HFB-QRPA \cite{Goriely2018} $M1$ strengths were employed together with the corresponding $E1$ model calculations. The resulting range of \textsc{talys} model uncertainties is displayed by broad beige bands in Figs. \ref{fig 10: ng CS and Rates} and \ref{fig 11: pg CS and Rates}. As default cross sections and rates, also shown in the figures by dashed-dotted lines, the \textsc{talys} software utilizes the CT plus Fermi gas NLD model, the Simple Modified Lorentzian (SMLO) models for the $E1$ and $M1$ strengths, and the global optical model potential by Koning and Delaroche (see Ref.~\cite{Koning2023} for more details regarding the used models).

No prior experimental information exists on the neutron capture cross section for the unstable $^{108}$In target; the result in Fig.~\ref{fig 10: ng CS and Rates}(a) provides the first experimental constraint on this quantity. The Oslo method data are, on average, 1.5 times higher than the \textsc{talys} default predictions, which is not unexpected given that both the default NLD and GSF inputs are lower in absolute values than the Oslo method results. Although the adopted $M1$ models vary considerably, their contribution to the experimental uncertainty band is minor; it is instead dominated by the uncertainties in the experimental NLD and the $E1$ GSF. Similarly, the contribution of the uncertainty due to the $M1$ models to the \textsc{talys} band does not exceed 3-4\%. Variation of the NLD models with the GSF and the optical potential fixed produces uncertainties from 10\% to approximately 20\% of the total \textsc{talys} uncertainty in the shown energy range. The uncertainty in the GSF model appears to be the largest contributor to the \textsc{talys} model uncertainty. With the NLD and the optical model potential held fixed, variation of the GSF models leads to uncertainties ranging from 30\% to 50\% of the band shown in Fig.~\ref{fig 10: ng CS and Rates}(a). Finally, similar calculations with the NLD and GSF models fixed show that the uncertainty due to the optical potential reaches its maximum of approximately 30\% at low energies and decreases to a few percent toward neutron energies of 1 MeV.

In the neutron energy range shown in Fig.~\ref{fig 10: ng CS and Rates}(a), the CT and the BSFG models combined with the temperature-dependent
relativistic mean field plus QRPA $E1$ calculations yield the lowest neutron-capture cross section (bottom of the \textsc{talys} uncertainty range). Unsurprisingly, the microscopic Skyrme-HF-BCS plus statistical calculations, yielding the highest NLD values, combined with the simple Lorentzian for the $E1$ strength, significantly overshooting the experimental data at low energies, result in the highest ($n,\gamma$) cross section. It is noteworthy that the combination of the same NLD model with the hybrid $E1$ model \cite{Goriely98} yields the best agreement with the experimentally constrained cross section.

The experimental ($n,\gamma$) reaction rate is shown in Fig.~\ref{fig 10: ng CS and Rates}(b) together with the predictions from the JINA REACLIB \cite{JINA} and BRUSLIB libraries \cite{BRUSLIB}. The latter employs the Skyrme-HFB plus combinatorial NLD with the Gogny-HFB plus QRPA GSF and reproduces the experimental data relatively well. The JINA REACLIB rate (model ths8, version 6), despite having been shown to perform well for Sn isotopes \cite{Markova2024}, disagrees notably with both BRUSLIB and the experiment. This discrepancy may be a common feature for neutron-deficient and the lightest stable isotopes \cite{Larsen2013_Cd, Bell2025} and should be taken into consideration for $p$-process calculations with JINA REACLIB reaction rates.

The radiative proton capture on stable $^{108}$Cd has been previously studied and presented in Refs.~\cite{Skakun1992,Gyürky2007,Olivas-Gomez2020, Saha2025}, and the comparison of the extracted ($p,\gamma$) cross sections with the present results is shown in Fig.~\ref{fig 11: pg CS and Rates}(a). The work by Olivas-Gomez \textit{et al.} \cite{Olivas-Gomez2020} specifically focuses on constraining the NLD and the GSF models with experimental proton capture cross sections on $^{108,110}$Cd, aiming to improve the description of the ($\gamma, n$) and $(\gamma, p)$ rates for the potential $p$-process branching point $^{111}$In. A similar objective to constrain the nuclear input models was also pursued in the most recent study by Saha \textit{et al.} \cite{Saha2025}. Within the uncertainties, our data are fully consistent with the more recent analyses by Gyürky \textit{et al.} \cite{Gyürky2007}, Olivas-Gomez \textit{et al.} \cite{Olivas-Gomez2020}, and  Saha \textit{et al.} \cite{Saha2025} over a wide range of energies from approximately 3 MeV to 7 MeV and with the older data by Skakun \textit{et al.} \cite{Skakun1992} for 7 MeV and higher. This serves as another benchmark for the Oslo method inputs, supporting the assumptions made in the normalization procedure. Similarly to the neutron-capture case, the \textsc{talys} default calculations tend to slightly underestimate the experimental cross sections. The microscopic Skyrme-HF-BCS plus statistical NLD from Ref.~\cite{Demetriou2001} and the Skyrme-HFB-QRPA GSF from Ref.~\cite{Xu2021} were found to be an optimal combination yielding the best agreement with the Oslo data. The latter model, however, underestimates the IVGDR peak quite notably compared to other models available in \textsc{talys}. 

In general, our results support the recommendation made in Ref.~\cite{Olivas-Gomez2020}, proposing the Skyrme-HF-BCS NLD by Demetriou \textit{et al.}~\cite{Demetriou2001} as a suitable model for cross section calculations in the vicinity of $A\approx 109$ if no experimental data are available. It is important to note, however, that the NLD extracted in this work deviates from this model already at excitation energies of $\approx 6$ MeV.  The GSF based on the Skyrme-HF-BCS-QRPA calculations by Goriely and Khan \cite{Goriely2002}, recommended in the same work, does not yield the best overlap with our data when combined with the Skyrme-HF-BCS NLD model. Including an additional $E1$ enhancement at $\approx 7-8$ MeV, with a strength of about $0.5\%$ of the TRK sum rule as extracted in this work, would improve the agreement between the calculated cross section and the experimental data. The Skyrme-HFB NLD and the simple Lorentzian GSF were recommended in the recent work by Saha \textit{et al.} in combination with the newly calculated optical potential for p-nuclei (see Ref.~\cite{Saha2025} for more details) to reproduce the experimental cross sections. In the present work, however, these models are shown to exhibit significant deviations from the Oslo method NLD and GSF. It is interesting to note that the $M1$ response, despite a great variability in the $M1$ strength distribution between different models \cite{Koning2023}, is a rather minor contributor to the uncertainty of both the $(n,\gamma)$ and $(p,\gamma)$ cross sections, compared to the $E1$ strength. At proton energies below 3 MeV, the optical model potential dominates the uncertainty band,  whereas at higher energies the cross section becomes considerably more sensitive to the choice of the NLD and GSF.

Finally, the ($p, \gamma$) rates were estimated using the Oslo method inputs; the results of the calculations are shown in Fig.~\ref{fig 11: pg CS and Rates}(b) together with the \textsc{talys} default rate and model uncertainty, as well as the rates from the JINA REACLIB and BRUSLIB libraries. In this case, the JINA REACLIB rate is more consistent with BRUSLIB, but both of them deviate from the experimentally constrained rate at temperatures of $6-9$ GK. The $p$-process sensitivity to the $^{109}$In($\gamma,p$)$^{108}$Cd and $^{109}$In($\gamma,n)$$^{108}$In rates, estimated with the detailed balance theorem from the experimentally constrained $^{108}$Cd($p,\gamma$)$^{109}$In and $^{108}$In($n,\gamma)$$^{109}$In rates, appears to be marginal. This is to be expected, considering that production of the $^{106}$Cd and $^{108}$Cd isotopes is dominated by photodissociation of heavier Cd isotopes. The results of the recent study by Saha \textit{et al.} and the earlier work by Olivas-Gomez \textit{et al.} highlight a persistent challenge in statistical-model calculations of reaction rates for the p process, namely that direct measurements of ($p,\gamma$) cross sections alone do not fully constrain NLD and GSF models for p-process nuclei. The impact of the new experimental results is therefore primarily indirect: in combination with direct ($p,\gamma$) measurements, the newly extracted NLD and GSF offer a complementary perspective on the performance of theoretical models and on reducing the associated parameter uncertainties, as well as on the low-lying $E1$ enhancement and its magnitude in neutron-deficient nuclei in the Sn mass region.


\section{\label{sec 5: Conclusion}Conclusions and outlook}

In this work, the NLD and the GSF of the neutron-deficient $^{109}$In isotope were extracted with the Oslo and the shape methods from $p-\gamma$ coincidence data. The NLD appears to follow a clear BSFG behavior at excitation energies above $\approx$ 5 MeV, well below the neutron threshold. It is at variance with the majority of the theoretical model predictions, underestimating the experimental data at all energies. The GSF agrees well with the results for the neighboring Sn and Cd isotopes at low energies but displays no clear $E1$ enhancement around $7-8$ MeV, observed earlier in Pd, Cd, and Sn nuclei. In this case, the low-lying $E1$ strength on top of the IVGDR was estimated to be $\approx 0.5$\% of the TRK sum rule, which is considerably lower than in the previously studied neighboring nuclei. The RQTBA calculations, performed to provide deeper insight into the qualitative change of the $E1$ strength from $^{112}$Sn to $^{109}$In, indicate a notable suppression of the neutron contribution to the excitations in $^{109}$In below the neutron threshold compared to $^{112}$Sn.

The cross-sections and rates of the $^{108}$Cd($p,\gamma$)$^{109}$In and $^{108}$In($n,\gamma)$$^{109}$In reactions were estimated within the Hauser-Feshbach framework using the new experimental NLD and GSF as inputs, thus providing important constraints on the presently available model uncertainties. Our results are in excellent agreement with earlier ($p,\gamma$) cross section measurements. The ($n, \gamma$) reaction rate from the JINA REACLIB library, despite its good performance for the neighboring heavier nuclei, deviates significantly from the BRUSLIB predictions and the experimental result. This point could be relevant for future $p$-process calculations utilizing this library.

We hope that the new information on the low-lying $E1$ strength in $^{109}$In reported in this work will encourage further research on other neutron-deficient nuclei in this mass region, as well as further improvements of theoretical NLD and GSF models for more accurate $p$-process simulations.

\begin{acknowledgments}
The authors express their thanks to P.~A.~Sobas, V. Modamio and J.~C.~Wikne at the Oslo Cyclotron Laboratory for operating the cyclotron and providing excellent experimental conditions. The authors are also thankful to D. Gjestvang for her contribution to the
experiment. P.v.N.-C. thanks the nuclear physics group at the University of Oslo for their kind hospitality during the stay where parts of the present work were done. This work was funded by the Research Council of Norway through its grants to the Norwegian Nuclear Research Centre (Project No.~341985) and the nuclear physics research group at the University of Oslo (Project No.~245882). 
Funding from the Research Council of Norway is gratefully acknowledged by A.~C.~L. (Project No.~316116), E. M. (Project No. 310094), S. S. (Project No. 263030),  L.~T.~B., A.~G., S.~S., W.~P. (Project No. 325714).  P.v.N.-C. acknowledges support by the Deutsche Forschungsgemeinschaft (DFG, German Research Foundation) under Grant No. SFB 1245 (Project ID No. 279384907). The work of E.L. was supported by the GANIL Visitor Program, US-NSF Grant PHY-2209376, and US-NSF Grant PHY-2515056. S.G. acknowledges support from the European Union (ChETEC-INFRA, project No. 101008324). This work was supported by the F.R.S.-FNRS under Grant No. IISN 4.4502.19. S.G. is a senior F.R.S.-FNRS research associate.

\end{acknowledgments}


\bibliography{In_2025}
\end{document}